\documentclass{article}
\usepackage{slashed,verbatim,subfigure}
\usepackage[numbers,sort&compress]{natbib}
\usepackage{amsmath}
\usepackage{amssymb}
\usepackage{appendix}
\usepackage{graphicx}
\usepackage{dcolumn}
\usepackage{bm}
\usepackage[colorlinks=true,linktocpage=true,
linkcolor=blue,citecolor=blue]{hyperref}
\usepackage[a4paper]{geometry}
\newcommand{\be}{\begin{eqnarray}}
\newcommand{\ee}{\end{eqnarray}}
\begin{document}
\large
\title{\bf{Viscous properties of hot and dense QCD matter in the presence of a magnetic field}}
\author{Shubhalaxmi Rath\footnote{srath@ph.iitr.ac.in}~~and~~Binoy Krishna 
Patra\footnote{binoy@ph.iitr.ac.in}\vspace{0.03in} \\ 
Department of Physics, Indian Institute of Technology Roorkee, Roorkee 247667, India}
\date{}
\maketitle
\begin{abstract}
We have studied the effect of strong magnetic field on the viscous properties of hot QCD matter at finite chemical potential by calculating the shear viscosity ($\eta$) and the bulk viscosity ($\zeta$). The viscosities have been calculated using the relativistic Boltzmann transport equation within the relaxation time approximation. The interactions among partons are incorporated through their quasiparticle masses at finite temperature, strong magnetic field and finite chemical potential. From this study, one can understand the influence of strong magnetic field and the influence of chemical potential on the sound attenuation through the Prandtl number (Pl), on the nature of the flow by the Reynolds number (Rl), and on the relative behavior between the shear viscosity and the bulk viscosity through the ratio $\zeta/\eta$. We have observed that, both shear and bulk viscosities get increased in the presence of a strong magnetic field and the additional presence of chemical potential further enhances their magnitudes. With the increase of temperature, $\eta$ increases for the medium in the presence of a strong magnetic field as well as for the isotropic medium in the absence of magnetic field, whereas $\zeta$ is found to decrease with the temperature, contrary to its increase in the absence of magnetic field. We have observed that, the Prandtl number gets increased in the presence of strong magnetic field and finite chemical potential as compared to that in the isotropic medium, but it always remains larger than unity, thus instead of the thermal diffusion, the momentum diffusion largely affects the sound attenuation in the medium and this is more vigorous in the presence of both strong magnetic field and finite chemical potential. However, the Reynolds number becomes lowered than unity in an ambience of strong magnetic field and even gets further decreased in an additional presence of chemical potential, thus it implies the dominance of kinematic viscosity over the characteristic length scale of the system. Finally, the ratio $\zeta/\eta$ is amplified to the value larger than unity, contrary to its value in the absence of magnetic field and chemical potential where it is less than unity, thus it is inferred that the bulk viscosity prevails over the shear viscosity for the hot and dense QCD matter in the presence of a strong magnetic field. 

\end{abstract}

\newpage

\section{Introduction}
The properties of the deconfined state of matter, {\em i.e.} the quark-gluon plasma (QGP) produced in the initial stages of ultrarelativistic heavy-ion collisions (URHICs) at Relativistic Heavy Ion Collider (RHIC) \cite{Arsene:NPA757'2005,Adams:NPA757'2005} and Large Hadron Collider (LHC) \cite{Carminati:JPG30'2004,Alessandro:JPG32'2006}, depend upon the initial conditions, such as the high temperatures and/or the strong magnetic fields and/or the finite chemical potentials. The condition of strong magnetic field arises in case of the noncentral collisions, where it has been observed that the strength of such magnetic field varies from $eB=m_{\pi}^2$ ($\simeq 10^{18}$ Gauss) at RHIC to 15 $m_{\pi}^2$ at LHC \cite{Skokov:IJMPA24'2009} and the condition of large chemical potential is expected to be evidenced in the Compressed Baryonic Matter (CBM) experiment at Facility for Antiproton and Ion Research (FAIR) \cite{Senger:CEJP10'2012}. The estimation for the lifetime of 
the strong magnetic field shows that it only exists for a small fraction of the lifetime of QGP, however, the electrical conductivity of QGP may significantly increase its lifetime \cite{Tuchin:AHEP2013'2013,Conductivities}. According to some observations, the quark chemical potential may reach approximately 100 MeV near the critical temperature (around 160 MeV) of phase transition \cite{P:JPG28'2002,Cleymans:JPG35'2008,Andronic:NPA837'2010} and in the presence of a strong magnetic field, it rises up to 200 MeV \cite{Fukushima:PRL117'2016}. Thus, various properties of the hot medium of quarks and gluons are prone to be affected by the existence of strong magnetic field and finite chemical potential. Different observations have already been made in discerning the effect of strong magnetic field on the properties 
of hot QCD matter, such as the thermodynamic and magnetic properties \cite{Bandyopadhyay:PRD100'2019,Rath:JHEP1712'2017,Expansion,Karmakar:PRD99'2019}, the chiral magnetic effect \cite{Fukushima:PRD78'2008,Kharzeev:NPA803'2008}, the axial magnetic effect \cite{Braguta:PRD89'2014,Chernodub:PRB89'2014}, the dilepton production from QGP \cite{Tuchin:PRC88'2013,Mamo:JHEP1308'2013} etc. 

In an ambience of external magnetic field, the dispersion relation of $f$th flavor of quark with absolute charge $|q_f|$ and mass $m_f$ is written as $\omega_{f,n}=\sqrt{p_L^2+2n|q_fB|+m_f^2}$, where $p_L$ is the longitudinal component of momentum (with respect to the direction of magnetic field) and the momentum along the transverse direction ($p_T$) is quantized in terms of the Landau levels ($n$). The presence of strong magnetic field makes the charged particles or the quarks to reside only in the lowest Landau level (LLL) and their motion to be along one spatial dimension, {\em i.e.} along the direction of magnetic field, because the charged particles can not jump to the higher Landau levels due to very high energy gap $\sim\mathcal{O}(\sqrt{|q_fB|})$. Thus, in the strong magnetic field (SMF) limit ($|q_fB| \gg T^2$ and $|q_fB| \gg m_f^2$), $p_L$ is much greater than $p_T$, for which an anisotropy is created in the momentum space. Unlike the quarks, the gluons are not directly affected by the magnetic field because they are electrically uncharged particles, however, they can be indirectly affected by the magnetic field through their thermal masses. According to some observation \cite{Fukushima:JHEP2004'2020}, the effect of magnetic field on the longitudinal electrical conductivity is predominant in the presence of finite chemical potential. So, the properties of transport coefficients might be robustly influenced by the emergence of strong magnetic field and finite chemical potential. In this process, we have recently studied the charge and thermal transport properties by calculating the electrical conductivity ($\sigma_{\rm el}$) and the thermal conductivity ($\kappa$) in the presence of both strong magnetic field and chemical potential and also observed their applications to understand the effects of strong magnetic field and chemical potential on the local equilibrium by the Knudsen number ($\Omega$) and on the Lorenz number ($L$) in the Wiedemann-Franz law \cite{Rath:EPJC80'2020}. Our aim in this work is to investigate the viscous properties of the hot QCD matter by calculating the shear viscosity ($\eta$) and the bulk viscosity ($\zeta$) in the abovementioned environment of strong magnetic field and finite but small quark chemical potential. Then we intend to observe some applications in the similar environment, such as the sound attenuation through the Prandtl number (Pl), the nature of the flow by the Reynolds number (Rl), and the relative behavior between the shear viscosity and the bulk viscosity through the ratio $\zeta/\eta$. 

A system which is slightly out of equilibrium can possess finite shear and bulk viscosities. In the hydrodynamic description of QGP, two of the important quantities are the shear viscosity and the bulk viscosity, out of which shear viscosity conducts the momentum transfer in the presence of inhomogeneity of fluid velocity, whereas the bulk viscosity delineates the change of local pressure due to either expansion or contraction of fluid.  In the study of phase transition from hadronic matter to quark and gluon matter, the values of the viscosities are helpful in determining the location of phase transition, where the shear viscosity is a minimum and the bulk viscosity is a maximum \cite{Csernai:PRL97'2006}. In hydrodynamic simulations, different observables, such as the elliptic flow coefficient and the hadron transverse momentum spectrum are largely influenced by the shear and bulk viscosities \cite{Song:PLB658'2008,Denicol:JPG37'2010,Dusling:PRC85'2012,Noronha-Hostler:PRC90'2014}. Their values and properties also provide the information on how far the system appears from an ideal hydrodynamics. In this regard, a variety of calculations on shear and bulk viscosities have been done for a medium consisting of quarks and gluons at high temperatures by implementing the perturbation theory \cite{Arnold:JHEP'2000'2003,Arnold:PRD74'2006,Hidaka:PRD78'2008}, the kinetic theory \cite{Danielewicz:PRD31'1985,Sasaki:PRC79'2009,Thakur:PRD95'2017} etc. The presence of magnetic field breaks the rotational symmetry, therefore the viscous stress tensor is described by five shear viscous coefficients and two bulk viscous coefficients \cite{Lifshitz:BOOK'1981,Tuchin:JPG39'2012,Critelli:PRD90'2014,
Hernandez:JHEP05'2017,Hattori:PRD96'2017,Chen:PRD101'2020}. However, in the strong magnetic field limit, only the longitudinal components (along the direction of magnetic field) of shear and bulk viscosities exist and these are contributed by the lowest Landau level (LLL) quarks/antiquarks \cite{Lifshitz:BOOK'1981,Tuchin:JPG39'2012,Hattori:PRD96'2017}, so, the viscosities become highly anisotropic. The relativistic anisotropic viscosities were first introduced in references \cite{Huang:PRD81'2010,Huang:AP326'2011} and a kinetic formalism of these transport coefficients was given in references \cite{Denicol:PRD98'2018,Denicol:PRD99'2019}. In the presence of the magnetic field, the shear and bulk viscosities have been previously determined using various approaches and techniques, {\em viz.} the correlator technique using Kubo formula \cite{Nam:PRD87'2013,Hattori:PRD96'2017}, the perturbative QCD approach in weak magnetic field \cite{Li:PRD97'2018}, the Chapman-Enskog method with effective fugacity approach \cite{Kurian:EPJC79'2019} and the holographic setup \cite{Rebhan:PRL108'2012,Jain:JHEP10'2015,Finazzo:PRD94'2016}. In this work, we are going to calculate these viscosities by using the kinetic theory approach in the relaxation time approximation, where we have considered the interactions among particles through their effective masses in the quasiparticle model at strong magnetic field and finite chemical potential. The study of the effects of strong magnetic field and chemical potential on the transport coefficients is important, as in the ultrarelativistic heavy ion collisions, the rapid evolution of the matter is governed by the dissipative effects via the shear viscosity and the bulk viscosity and is expected to be modified by the strong magnetic field and chemical potential. 

We intend to further study the impacts of strong magnetic field and finite chemical potential on the Prandtl number (Pl=$\frac{\eta C_p}{\rho \kappa}$, where $C_p$ denotes the specific heat at constant pressure, $\rho$ is the mass density and $\kappa$ represents the thermal conductivity), on the Reynolds number (Rl=$\frac{Lv\rho}{\eta}$, where $L$ and $v$ denote the characteristic length and velocity of the flow, respectively), and on the ratio of bulk viscosity to shear viscosity ($\zeta/\eta$). The Prandtl number signifies the relative importance of the momentum diffusion and the thermal diffusion on the sound attenuation in a medium. Using kinetic theory, the Prandtl number has been calculated for a strongly coupled liquid helium \cite{T:RPP72'2009}, where its value is found to be around 2.5, and for a nonrelativistic conformal holographic fluid, Pl is estimated to be 1.0 \cite{T:RPP72'2009,Rangamani:JHEP01'2009}. The Prandtl number is $\frac{2}{3}$ for a dilute atomic Fermi gas at high temperatures \cite{Braby:PRA82'2010}. The magnitude of the Reynolds number gives the knowledge about the type of flow, {\em i.e.} the flow is laminar when ${\rm Rl} \leq 1$ and it is turbulent when ${\rm Rl} \gg 1$. The Reynolds number of the quark matter is calculated to be about 10 using the Kubo formula and NJL model \cite{Fukutome:PTP119'2008}, according to the (3+1) dimensional fluid dynamical model, the value of the Rl for QGP lies in the range 3-10 \cite{Csernai:PRC85'2012} and the holographic setup reports the upper value of Rl as approximately 20 \cite{McInnes:NPB921'2017}. In neither of the abovementioned approaches Rl is much larger than 1, so the QGP medium can be considered as a viscous medium with laminar flow. The properties of the ratio $\zeta/\eta$ has also been studied previously for different systems using different approaches. For example, $\zeta/\eta$ is studied for an interacting scalar field in ref. \cite{Horsley:NPB280'1987}, for a hot QCD medium at perturbative limit in ref. \cite{Arnold:PRD74'2006}, for a strongly coupled gauge theory plasma in ref. \cite{Buchel:PLB663'2008}, for a hot QCD medium using holographic model in references \cite{Buchel:NPB820'2009,U:JHEP0912'2009}, for a quasigluon plasma in ref. \cite{Bluhm:PLB709'2012}, for hadron gas in references \cite{Prakash:PR227'1993,Davesne:PRC53'1996,Chen:PRC79'2009} and for a hot QCD medium using the Chapman-Enskog method in ref. \cite{Mitra:PRD96'2017}. In these systems, the ratio $\zeta/\eta$ has different values due to different system dynamics at different coupling regimes. 

The present work is organized as follows. In section 2, we have first studied the viscous properties by calculating the shear and bulk viscosities for a hot QCD medium at finite chemical potential in the absence of magnetic field and then proceeded to calculate the same in the presence of a strong magnetic field. Then, we have studied some applications of the shear and bulk viscous properties, {\em viz.}, the Prandtl number, the Reynolds number and the relative behavior between the shear viscosity and the bulk viscosity in section 3. In section 4, we have studied the quasiparticle model in the presence of both strong magnetic field and chemical potential. Section 5 contains the discussions on our results regarding shear and bulk viscosities, Prandtl number, Reynolds number and ratio of bulk viscosity to shear viscosity in quasiparticle description. Finally, we have concluded in section 6. 

\section{Shear and bulk viscous properties}
In this section, we are going to study the shear and bulk viscous properties of the QCD medium. It is possible to determine the shear and bulk viscosities using various approaches, {\em viz.} the relativistic Boltzmann transport equation in the relaxation time approximation \cite{Danielewicz:PRD31'1985,Heckmann:EPJA48'2012,Yasui:PRD96'2017}, the lattice simulation \cite{Astrakhantsev:JHEP1704'2017,Astrakhantsev:PRD98'2018}, the correlator technique using Green-Kubo formula \cite{Basagoiti:PRD66'2002,Kharzeev:JHEP0809'2008,Moore:JHEP0809'2008,Plumari:PRC86'2012}, the molecular dynamics simulation \cite{Gelman:PRC74'2006} etc. In our work, we use the relativistic Boltzmann transport equation to calculate the shear and bulk viscosities within the relaxation time approximation for a dense QCD medium in the absence of magnetic field and for a dense QCD medium in the presence of a strong magnetic field in subsections 2.1 and 2.2, respectively. 

\subsection{Hot and dense QCD medium in the absence of magnetic field}
When the medium deviates slightly from the equilibrium, the 
energy-momentum tensor also becomes shifted as
\begin{eqnarray}
\Delta T^{\mu\nu}=T^{\mu\nu}-T_{(0)}^{\mu\nu}
~,\end{eqnarray}
where $T_{(0)}^{\mu\nu}$ is the energy-momentum tensor 
in local equilibrium and $T^{\mu\nu}$ denotes the total 
energy-momentum tensor in a nonequilibrium medium. $T^{\mu\nu}$ is defined as
\be
T^{\mu\nu}=\int\frac{d^3{\rm p}}{(2\pi)^3}p^\mu p^\nu \left[\sum_f g_f\frac{\left(f_f+\bar{f}_f\right)}{{\omega_f}}+g_g\frac{f_g}{\omega_g}\right]
.\ee
Similarly, $\Delta T^{\mu\nu}$ is written as
\be\label{em1}
\Delta T^{\mu\nu}=\int\frac{d^3{\rm p}}{(2\pi)^3}p^\mu p^\nu \left[\sum_f g_f\frac{\left(\delta f_f+\delta \bar{f}_f\right)}{{\omega_f}}+g_g\frac{\delta f_g}{\omega_g}\right]
,\ee
where `$f$' is used for flavor index and it takes flavors $u$, $d$ and $s$. In eq. \eqref{em1}, $g_f$ and $\delta f_f$ ($\delta \bar{f_f}$) are the degeneracy factor and the infinitesimal change in the distribution function for the quark (antiquark) of $f$th flavor, respectively. Similarly, $g_g$ and $\delta f_g$ are the degeneracy factor and the infinitesimal change in the distribution function for the gluon, respectively. For a nonequilibrium system, the shear and bulk viscosities are finite, which are defined as the coefficients of the traceless and trace parts of the nonequilibrium contribution of the energy-momentum tensor, respectively. Thus the shear and bulk viscosities explain the momentum transports across and along the layer, respectively. 

The infinitesimal change in quark (antiquark) distribution function 
due to the action of an external force is defined as 
$\delta f_f=f_f-f_f^{\rm iso}$ ($\delta \bar{f}_f=\bar{f}_f-\bar{f}_f^{\rm iso}$), where $f_f^{\rm iso}$ ($\bar{f}_f^{\rm iso}$) represents the equilibrium distribution function in the isotropic medium for quark (antiquark) of $f$th flavor. The forms of $f_f^{\rm iso}$ and $\bar{f}_f^{\rm iso}$ are given by
\be\label{Q.D.F.}
&&f_f^{\rm iso}=\frac{1}{e^{\beta \left(u^\alpha p_\alpha-\mu_f\right)}+1} ~, \\
&&\label{Q.D.F.1}\bar{f}_f^{\rm iso}=\frac{1}{e^{\beta \left(u^\alpha p_\alpha+\mu_f\right)}+1}
~,\ee
respectively, where $p_\alpha\equiv\left(\omega_f,\mathbf{p}\right)$ with 
$\omega_f=\sqrt{\mathbf{p}^2+m_f^2}$, $u^\alpha$ is the four-velocity 
of fluid, $T=\beta^{-1}$ and $\mu_f$ is the chemical potential of $f$th 
flavor. Similarly, the infinitesimal change in gluon distribution 
function is defined as $\delta f_g=f_g-f_g^{\rm iso}$, where 
$f_g^{\rm iso}$ is the equilibrium distribution function for gluon in the 
isotropic medium,
\be\label{G.D.F.}
f_g^{\rm iso}=\frac{1}{e^{\beta u^\alpha p_\alpha}-1}
~,\ee
with $p_\alpha\equiv\left(\omega_g,\mathbf{p}\right)$. The 
infinitesimal changes in the distribution functions for quarks, 
antiquarks and gluons can be determined from their respective 
relativistic Boltzmann transport equations in the relaxation-time approximation: 
\begin{eqnarray}
&&p^\mu\partial_\mu f_f(x,p) = -\frac{p_\nu u^\nu}{\tau_f}\delta f_f(x,p), \\
&&p^\mu\partial_\mu \bar{f}_f(x,p) = -\frac{p_\nu u^\nu}{\tau_{\bar{f}}}\delta \bar{f}_f(x,p), \\
&&p^\mu\partial_\mu f_g(x,p) = -\frac{p_\nu u^\nu}{\tau_g}\delta f_g(x,p)
,\end{eqnarray}
where the forms of the relaxation times for quarks (antiquarks), $\tau_f$ ($\tau_{\bar{f}}$) and gluons, $\tau_g$ are given \cite{Hosoya:NPB250'1985} by
\begin{eqnarray}
&&\tau_{f(\bar{f})}=\frac{1}{5.1T\alpha_s^2\log\left(1/\alpha_s\right)\left[1+0.12 
(2N_f+1)\right]} ~, \label{Q.R.T.} \\
&&\tau_g=\frac{1}{22.5T\alpha_s^2\log\left(1/\alpha_s\right)\left[1+0.06N_f
\right]} \label{G.R.T.}
~,\end{eqnarray}
respectively. Substituting the values of $\delta f_f$, $\delta \bar{f}_f$ and 
$\delta f_g$ in eq. \eqref{em1}, we get
\be\label{em2}
\Delta T^{\mu\nu}=-\int\frac{d^3{\rm p}}{(2\pi)^3}\frac{p^\mu p^\nu}{p_\nu u^\nu} \left[\sum_f g_f\frac{\left(\tau_f p^\mu\partial_\mu f_f+\tau_{\bar{f}} p^\mu\partial_\mu \bar{f}_f\right)}{\omega_f}+g_g\frac{\tau_g p^\mu\partial_\mu f_g}{\omega_g}\right]
.\ee
The derivative $\partial_\mu$ is written as $\partial_\mu=u_\mu D+\nabla_\mu$, where $D=u^\mu\partial_\mu$. In the local rest frame, the 
flow velocity and the temperature are treated as the functions of spatial and temporal 
coordinates, thus, it is possible to expand the distribution functions in terms 
of the gradients of flow velocity and temperature. The partial 
derivatives of the isotropic quark, antiquark and gluon distribution 
functions are determined as
\begin{eqnarray}
\partial_\mu f_f^{\rm iso}=\frac{f_f^{\rm iso}(1-f_f^{\rm iso})}{T}\left[u_\alpha p^\alpha u_\mu\frac{DT}{T}+u_\alpha p^\alpha\frac{\nabla_\mu T}{T}-u_\mu p^\alpha Du_\alpha-p^\alpha\nabla_\mu u_\alpha+T\partial_\mu\left(\frac{\mu_f}{T}\right)\right]
,\end{eqnarray}
\begin{eqnarray}
\partial_\mu \bar{f}_f^{\rm iso}=\frac{\bar{f}_f^{\rm iso}(1-\bar{f}_f^{\rm iso})}{T}\left[u_\alpha p^\alpha u_\mu\frac{DT}{T}+u_\alpha p^\alpha\frac{\nabla_\mu T}{T}-u_\mu p^\alpha Du_\alpha-p^\alpha\nabla_\mu u_\alpha-T\partial_\mu\left(\frac{\mu_f}{T}\right)\right]
,\end{eqnarray}
\begin{eqnarray}
\partial_\mu f_g^{\rm iso}=\frac{f_g^{\rm iso}(1+f_g^{\rm iso})}{T}\left[u_\alpha p^\alpha u_\mu\frac{DT}{T}+u_\alpha p^\alpha\frac{\nabla_\mu T}{T}-u_\mu p^\alpha Du_\alpha-p^\alpha\nabla_\mu u_\alpha\right]
,\end{eqnarray}
respectively. Substituting the values of $\partial_\mu f_f^{\rm iso}$, $\partial_\mu \bar{f}_f^{\rm iso}$ and $\partial_\mu f_g^{\rm iso}$ in eq. \eqref{em2} and then using $\frac{DT}{T}=-\left(\frac{\partial P}{\partial \varepsilon}\right)\nabla_\alpha u^\alpha$ and $Du_\alpha=\frac{\nabla_\alpha P}{\varepsilon+P}$ from the energy-momentum conservation, we have
\be\label{em3}
\nonumber\Delta T^{\mu\nu} &=& \sum_f g_f\int\frac{d^3{\rm p}}{(2\pi)^3}\frac{p^\mu p^\nu}{\omega_f T}\left[\tau_f ~ f_f^{\rm iso}(1-f_f^{\rm iso})\left\lbrace\omega_f\left(\frac{\partial P}{\partial \varepsilon}\right)\nabla_\alpha u^\alpha+p^\alpha\left(\frac{\nabla_\alpha P}{\varepsilon+P}-\frac{\nabla_\alpha T}{T}\right)\right.\right. \\ && \left.\left.\nonumber-\frac{Tp^\alpha}{\omega_f}\partial_\alpha\left(\frac{\mu_f}{T}\right)+\frac{p^\alpha p^\beta}{\omega_f}\nabla_\alpha u_\beta\right\rbrace+\tau_{\bar{f}} ~ \bar{f}_f^{\rm iso}(1-\bar{f}_f^{\rm iso})\left\lbrace\omega_f\left(\frac{\partial P}{\partial \varepsilon}\right)\nabla_\alpha u^\alpha\right.\right. \\ && \left.\left.\nonumber+p^\alpha\left(\frac{\nabla_\alpha P}{\varepsilon+P}-\frac{\nabla_\alpha T}{T}\right)+\frac{Tp^\alpha}{\omega_f}\partial_\alpha\left(\frac{\mu_f}{T}\right)+\frac{p^\alpha p^\beta}{\omega_f}\nabla_\alpha u_\beta\right\rbrace\right] \\ && \nonumber+g_g\int\frac{d^3{\rm p}}{(2\pi)^3}\frac{p^\mu p^\nu}{\omega_g T} ~ \tau_g ~ f_g^{\rm iso}(1+f_g^{\rm iso})\left[\omega_g\left(\frac{\partial P}{\partial \varepsilon}\right)\nabla_\alpha u^\alpha\right. \\ && \left.+p^\alpha\left(\frac{\nabla_\alpha P}{\varepsilon+P}-\frac{\nabla_\alpha T}{T}\right)+\frac{p^\alpha p^\beta}{\omega_g}\nabla_\alpha u_\beta\right]
.\ee
From the energy-momentum tensor, the pressure and the energy density are respectively obtained as $P=-\Delta_{\mu\nu}T^{\mu\nu}/3$ and 
$\varepsilon=u_\mu T^{\mu\nu}u_\nu$, where $\Delta_{\mu\nu}=g_{\mu\nu}-u_\mu u_\nu$ represents the projection tensor. In order to define the shear and bulk viscosities, one requires the nonzero velocity gradient. We note that, $u^\mu$ denotes the velocity of the baryon number flow in the Eckart frame, whereas it denotes the velocity of the energy flow in the Landau-Lifshitz frame. Thus, an arbitrariness is created while defining the velocity $u^\mu$, which can be circumvented by imposing the ``condition of fit" in the local rest frame, {\em i.e.} $\Delta T^{00}=0$ \cite{Albright:PRC93'2016}. So, only the space-space component of $\Delta T^{\mu\nu}$ is finite and proportional to the velocity gradient. From eq. \eqref{em3}, the space-space component of $\Delta T^{\mu\nu}$ is written as
\be\label{em4}
\nonumber\Delta T^{ij} &=& \sum_f g_f\int\frac{d^3{\rm p}}{(2\pi)^3}\frac{p^i p^j}{\omega_f T}\left[\tau_f ~ f_f^{\rm iso}(1-f_f^{\rm iso})\left\lbrace-\frac{p^kp^l}{2\omega_f}W_{kl}+\left(\omega_f\left(\frac{\partial P}{\partial \varepsilon}\right)-\frac{\rm p^2}{3\omega_f}\right)\partial_l u^l\right.\right. \\ && \left.\left.\nonumber -\frac{Tp^k}{\omega_f}\partial_k\left(\frac{\mu_f}{T}\right)+p^k\left(\frac{\partial_k P}{\varepsilon+P}-\frac{\partial_k T}{T} \right)\right\rbrace+\tau_{\bar{f}} ~ \bar{f}_f^{\rm iso}(1-\bar{f}_f^{\rm iso})\left\lbrace-\frac{p^kp^l}{2\omega_f}W_{kl}\right.\right. \\ && \left.\left.\nonumber +\left(\omega_f\left(\frac{\partial P}{\partial \varepsilon}\right)-\frac{\rm p^2}{3\omega_f}\right)\partial_l u^l+\frac{Tp^k}{\omega_f}\partial_k\left(\frac{\mu_f}{T}\right)+p^k\left(\frac{\partial_k P}{\varepsilon+P}-\frac{\partial_k T}{T} \right)\right\rbrace\right] \\ && \nonumber +g_g\int\frac{d^3{\rm p}}{(2\pi)^3}\frac{p^i p^j}{\omega_g T} ~ \tau_g ~ f_g^{\rm iso}(1+f_g^{\rm iso})\left[-\frac{p^kp^l}{2\omega_g}W_{kl}+\left(\omega_g\left(\frac{\partial P}{\partial \varepsilon}\right)-\frac{\rm p^2}{3\omega_g}\right)\partial_l u^l\right. \\ && \left.+p^k\left(\frac{\partial_k P}{\varepsilon+P}-\frac{\partial_k T}{T} \right)\right]
,\ee
where we have used the following expressions,
\be
\partial_k u_l &=&-\frac{1}{2}W_{kl}-\frac{1}{3}\delta_{kl}\partial_j u^j, \\
W_{kl} &=& \partial_k u_l+\partial_l u_k-\frac{2}{3}\delta_{kl}\partial_j u^j
.\ee

The shear and bulk viscosities are defined as the coefficients of the traceless and trace parts of the nonequilibrium part of energy-momentum tensor, respectively. For small deviation of the system from its equilibrium, the space-space component of the nonequilibrium part of energy-momentum tensor in a first order theory is written \cite{Lifshitz:BOOK'1981,Hosoya:NPB250'1985,Landau:BOOK'1987} as
\be\label{definition}
\Delta T^{ij}=-\eta W^{ij}-\zeta\delta^{ij}\partial_l u^l
.\ee
After comparing equations \eqref{em4} and \eqref{definition}, we get the shear viscosity and the bulk viscosity for an isotropic dense QCD medium in the absence of magnetic field as
\begin{eqnarray}\label{iso.eta}
\nonumber\eta^{\rm iso} &=& \frac{\beta}{30\pi^2}\sum_f g_f \int d{\rm p}~\frac{{\rm p}^6}{\omega_f^2}\left[\tau_f ~ f_f^{\rm iso}(1-f_f^{\rm iso})+\tau_{\bar{f}} ~ \bar{f}_f^{\rm iso}(1-\bar{f}_f^{\rm iso})\right] \\ && +\frac{\beta}{30\pi^2} g_g \int d{\rm p}~\frac{{\rm p}^6}{\omega_g^2} ~ \tau_g ~ f_g^{\rm iso}(1+f_g^{\rm iso})
~,\end{eqnarray}
\begin{eqnarray}\label{iso.zeta1}
\nonumber\zeta^{\rm iso} &=& \frac{1}{3}\sum_f g_f \int\frac{d^3{\rm p}}{(2\pi)^3}~\frac{{\rm p}^2}{\omega_f}\left[f_f^{\rm iso}(1-f_f^{\rm iso})A_f+\bar{f}_f^{\rm iso}(1-\bar{f}_f^{\rm iso})\bar{A}_f\right] \\ && +\frac{1}{3}g_g \int\frac{d^3{\rm p}}{(2\pi)^3}~\frac{{\rm p}^2}{\omega_g} ~ f_g^{\rm iso}(1+f_g^{\rm iso})A_g
~.\end{eqnarray}
The factors $A_f$, $\bar{A}_f$ and $A_g$ in eq. \eqref{iso.zeta1} are respectively given by
\begin{eqnarray}
&&A_f=\frac{\tau_f}{3T}\left[\frac{{\rm p}^2}{\omega_f}-3\left(\frac{\partial P}{\partial \varepsilon}\right)\omega_f\right],\label{Ai} \\ 
&&\bar{A}_f=\frac{\tau_{\bar{f}}}{3T}\left[\frac{{\rm p}^2}{\omega_f}-3\left(\frac{\partial P}{\partial \varepsilon}\right)\omega_f\right],\label{Ai.1} \\ 
&&A_g=\frac{\tau_g}{3T}\left[\frac{{\rm p}^2}{\omega_g}-3\left(\frac{\partial P}{\partial \varepsilon}\right)\omega_g\right]\label{Ag}
.\end{eqnarray}
In the local rest frame, to satisfy the Landau-Lifshitz condition ($\Delta T^{00}=0$), the factors $A_f$, $\bar{A}_f$ and $A_g$ need to be replaced as $A_f\rightarrow A_f^\prime=A_f-b_f\omega_f$, $\bar{A}_f\rightarrow \bar{A}_f^\prime=\bar{A}_f-\bar{b}_f\omega_f$ and $A_g\rightarrow A_g^\prime=A_g-b_g\omega_g$. From eq. \eqref{em3}, the Landau-Lifshitz conditions for terms $A_f$, $\bar{A}_f$ and $A_g$ are written as
\begin{eqnarray}
&&\sum_f g_f\int\frac{d^3{\rm p}}{(2\pi)^3}~\omega_f f_f^{\rm iso}(1-f_f^{\rm iso})\left(A_f-b_f\omega_f\right)=0 \label{A_i} ~,~ \\ 
&&\sum_f g_f\int\frac{d^3{\rm p}}{(2\pi)^3}~\omega_f \bar{f}_f^{\rm iso}(1-\bar{f}_f^{\rm iso})\left(\bar{A}_f-\bar{b}_f\omega_f\right)=0 \label{A_i.1} ~,~ \\ 
&&g_g\int\frac{d^3{\rm p}}{(2\pi)^3}~\omega_g f_g^{\rm iso}(1+f_g^{\rm iso})\left(A_g-b_g\omega_g\right)=0 \label{A_g}
~,\end{eqnarray}
respectively. The quantities $b_f$, $\bar{b}_f$ and $b_g$ are determined 
by solving equations \eqref{A_i}, \eqref{A_i.1} and \eqref{A_g}. Substituting  $A_f\rightarrow A_f^\prime$, $\bar{A}_f\rightarrow \bar{A}_f^\prime$ and 
$A_g\rightarrow A_g^\prime$ in eq. \eqref{iso.zeta1} and then simplifying, 
we obtain the bulk viscosity for an isotropic dense QCD medium in the absence 
of magnetic field as
\begin{eqnarray}\label{iso.zeta}
\nonumber\zeta^{\rm iso} &=& \frac{\beta}{18\pi^2}\sum_f g_f \int d{\rm p}~{\rm p}^2\left[\frac{{\rm p}^2}{\omega_f}-3\left(\frac{\partial P}{\partial \varepsilon}\right)\omega_f\right]^2\left[\tau_f ~ f_f^{\rm iso}(1-f_f^{\rm iso})+\tau_{\bar{f}} ~ \bar{f}_f^{\rm iso}(1-\bar{f}_f^{\rm iso})\right] \\ && +\frac{\beta}{18\pi^2}g_g\int d{\rm p}~{\rm p}^2\left[\frac{{\rm p}^2}{\omega_g}-3\left(\frac{\partial P}{\partial \varepsilon}\right)\omega_g\right]^2 \tau_g ~ f_g^{\rm iso}(1+f_g^{\rm iso})
~.\end{eqnarray}

\subsection{Hot and dense QCD medium in the presence of a strong magnetic field}
In an ambience of magnetic field, the quark momentum gets decomposed into the transverse ($p_T$) and longitudinal ($p_L$) components with respect to the direction of magnetic field (say, z or $3$-direction), where the transverse motion is quantized in terms of the Landau levels. Thus, the energy of the quark of $f$th flavor takes the following form,
\begin{eqnarray}\label{dispersion relation}
\omega_{f,n}(p_L)=\sqrt{p_L^2+2n\left|q_fB\right|+m_f^2}
~.\end{eqnarray}
Here, $n=0$, $1$, $2$, $\cdots$ denote different Landau levels. In the strong magnetic field (SMF) limit, the energy scale associated with the magnetic field is much greater than the energy scale associated with the temperature ($|q_fB| \gg T^2$). In this limit, quarks occupy only the lowest Landau level ($n=0$) and cannot move to the higher Landau levels, because the energy gap between the Landau levels is very large $\sim\mathcal{O}(\sqrt{|q_fB|})$. Thus, $p_T \ll p_L$ and this creates an anisotropy in the momentum space. The distribution functions for quark and antiquark are modified as
\be\label{A.D.F.(eB)}
&&f^B_f=\frac{1}{e^{\beta\left(u^\alpha\tilde{p}_\alpha-\mu_f\right)}+1} ~, \\
&&\label{A.D.F.(eB1)}\bar{f}^B_f=\frac{1}{e^{\beta(u^\alpha\tilde{p}_\alpha+\mu_f)}+1}
~,\ee
respectively. Here $\tilde{p}_\alpha\equiv\left(\omega_f,p_3\right)$, where $\omega_f$ in the SMF limit ($n=0$) is given by
\be
\omega_f=\sqrt{p_3^2+m_f^2}
~.\ee
However, the gluons being electrically uncharged particles are not influenced by the magnetic field. So, the gluon distribution function does not get modified even in the presence of a strong magnetic field. 

The total energy-momentum tensor ($\tilde{T}^{\mu\nu}=\tilde{T}_{(0)}^{\mu\nu}+\Delta\tilde{T}^{\mu\nu}$) for the nonequilibrium system in a strong magnetic field is written as
\be
\tilde{T}^{\mu\nu}=\sum_f \frac{g_f|q_fB|}{4\pi^2}\int d p_3 ~ \frac{\tilde{p}^\mu\tilde{p}^\nu}{\omega_f}\left(f_f+\bar{f}_f\right)
.\ee
In the above equation, the modified (integration) phase factor due to the strong magnetic field $\left(\int\frac{d^3{\rm p}}{(2\pi)^3}=\frac{|q_fB|}{2\pi}\int \frac{dp_3}{2\pi}\right)$ \cite{Gusynin:NPB462'1996,Bruckmann:PRD96'2017} has been used. Similarly, the dissipative part of the energy-momentum tensor is given by
\be\label{emb1}
\Delta\tilde{T}^{\mu\nu}=\sum_f \frac{g_f|q_fB|}{4\pi^2}\int d p_3 ~ \frac{\tilde{p}^\mu\tilde{p}^\nu}{\omega_f}\left(\delta f_f+\delta \bar{f}_f\right)
,\ee
where $\tilde{p}^\mu=(p^0,0,0,p^3)$ in SMF limit. The infinitesimal disturbances, $\delta f_f$ and $\delta \bar{f}_f$ can be obtained by solving the relativistic Boltzmann transport equations for quark and antiquark distribution functions in the relaxation time approximation in a strong magnetic field, 
\begin{eqnarray}
&&\tilde{p}^\mu\partial_\mu f_f(x,p) = -\frac{\tilde{p}_\nu u^\nu}{\tau^B_f}\delta f_f ~, \\
&&\tilde{p}^\mu\partial_\mu \bar{f}_f(x,p) = -\frac{\tilde{p}_\nu u^\nu}{\tau^B_{\bar{f}}}\delta \bar{f}_f
~.\end{eqnarray}
Here, the relaxation time, $\tau^B_{f(\bar{f})}$ in the strong magnetic field limit is given \cite{Hattori:PRD95'2017} by
\begin{eqnarray}
\tau^B_{f(\bar{f})}=\frac{\omega_f\left(e^{\beta\omega_f}-1\right)}{\alpha_sC_2m_f^2\left(e^{\beta\omega_f}+1\right)}\frac{1}{\int dp^\prime_3\frac{1}{\omega^\prime_f\left(e^{\beta\omega^\prime_f}+1\right)}}
~,\end{eqnarray}
where $C_2$ is the Casimir factor. Substituting the values of $\delta f_f$ and $\delta \bar{f}_f$ in eq. \eqref{emb1}, we get
\be\label{emb2}
\Delta\tilde{T}^{\mu\nu}=-\sum_f \frac{g_f|q_fB|}{4\pi^2}\int d p_3 ~ \frac{\tilde{p}^\mu\tilde{p}^\nu}{\tilde{p}_\nu u^\nu \omega_f}\left(\tau^B_f \tilde{p}^\mu\partial_\mu f_f+\tau^B_{\bar{f}} \tilde{p}^\mu\partial_\mu \bar{f}_f\right)
.\ee
The partial derivatives of the quark and antiquark distribution functions in the presence of a strong magnetic field are respectively calculated as
\begin{eqnarray}
\partial_\mu f_f^B=\frac{f_f^B(1-f_f^B)}{T}\left[u_\alpha \tilde{p}^\alpha u_\mu\frac{DT}{T}+u_\alpha \tilde{p}^\alpha\frac{\nabla_\mu T}{T}-u_\mu \tilde{p}^\alpha Du_\alpha-\tilde{p}^\alpha\nabla_\mu u_\alpha+T\partial_\mu\left(\frac{\mu_f}{T}\right)\right]
,\end{eqnarray}
\begin{eqnarray}
\partial_\mu \bar{f}_f^B=\frac{\bar{f}_f^B(1-\bar{f}_f^B)}{T}\left[u_\alpha \tilde{p}^\alpha u_\mu\frac{DT}{T}+u_\alpha \tilde{p}^\alpha\frac{\nabla_\mu T}{T}-u_\mu \tilde{p}^\alpha Du_\alpha-\tilde{p}^\alpha\nabla_\mu u_\alpha-T\partial_\mu\left(\frac{\mu_f}{T}\right)\right]
.\end{eqnarray}
After substituting the expressions of $\partial_\mu f_f^B$ and $\partial_\mu \bar{f}_f^B$ in eq. \eqref{emb2} and then simplifying, we get
\be\label{emb2.1}
\nonumber\Delta \tilde{T}^{\mu\nu} &=& \sum_f \frac{g_f|q_fB|}{4\pi^2}\int d p_3 ~ \frac{\tilde{p}^\mu \tilde{p}^\nu}{\omega_f T}\left[\tau_f ~ f_f^{\rm iso}(1-f_f^{\rm iso})\left\lbrace\omega_f\left(\frac{\partial P}{\partial \varepsilon}\right)\nabla_\alpha u^\alpha\right.\right. \\ && \left.\left.\nonumber+\tilde{p}^\alpha\left(\frac{\nabla_\alpha P}{\varepsilon+P}-\frac{\nabla_\alpha T}{T}\right)-\frac{T\tilde{p}^\alpha}{\omega_f}\partial_\alpha\left(\frac{\mu_f}{T}\right)+\frac{\tilde{p}^\alpha \tilde{p}^\beta}{\omega_f}\nabla_\alpha u_\beta\right\rbrace+\tau_{\bar{f}} ~ \bar{f}_f^{\rm iso}(1-\bar{f}_f^{\rm iso})\right. \\ && \left.\nonumber\times\left\lbrace\omega_f\left(\frac{\partial P}{\partial \varepsilon}\right)\nabla_\alpha u^\alpha+\tilde{p}^\alpha\left(\frac{\nabla_\alpha P}{\varepsilon+P}-\frac{\nabla_\alpha T}{T}\right)\right.\right. \\ && \left.\left.+\frac{T\tilde{p}^\alpha}{\omega_f}\partial_\alpha\left(\frac{\mu_f}{T}\right)+\frac{\tilde{p}^\alpha \tilde{p}^\beta}{\omega_f}\nabla_\alpha u_\beta\right\rbrace\right]
.\ee
The space-space or longitudinal component of $\Delta \tilde{T}^{\mu\nu}$ is written as
\be\label{emb3}
\nonumber\Delta \tilde{T}^{ij} &=& \sum_f \frac{g_f|q_fB|}{4\pi^2}\int d p_3 ~ \frac{\tilde{p}^i \tilde{p}^j}{\omega_f T}\left[\tau_f ~ f_f^{\rm iso}(1-f_f^{\rm iso})\left\lbrace-\frac{\tilde{p}^k\tilde{p}^l}{2\omega_f}W_{kl}+\left(\omega_f\left(\frac{\partial P}{\partial \varepsilon}\right)-\frac{p_3^2}{3\omega_f}\right)\partial_l u^l\right.\right. \\ && \left.\left.\nonumber -\frac{T\tilde{p}^k}{\omega_f}\partial_k\left(\frac{\mu_f}{T}\right)+\tilde{p}^k\left(\frac{\partial_k P}{\varepsilon+P}-\frac{\partial_k T}{T} \right)\right\rbrace+\tau_{\bar{f}} ~ \bar{f}_f^{\rm iso}(1-\bar{f}_f^{\rm iso})\left\lbrace-\frac{\tilde{p}^k\tilde{p}^l}{2\omega_f}W_{kl}\right.\right. \\ && \left.\left. +\left(\omega_f\left(\frac{\partial P}{\partial \varepsilon}\right)-\frac{p_3^2}{3\omega_f}\right)\partial_l u^l+\frac{T\tilde{p}^k}{\omega_f}\partial_k\left(\frac{\mu_f}{T}\right)+\tilde{p}^k\left(\frac{\partial_k P}{\varepsilon+P}-\frac{\partial_k T}{T} \right)\right\rbrace\right]
.\ee
In the strong magnetic field regime, the pressure and the energy density can be calculated from the energy-momentum tensor as $P=-\Delta^\parallel_{\mu\nu}\tilde{T}^{\mu\nu}$ and $\varepsilon=u_\mu\tilde{T}^{\mu\nu}u_\nu$, respectively, where the longitudinal projection tensor: $\Delta^\parallel_{\mu\nu}=g^\parallel_{\mu\nu}-u_\mu u_\nu$ with the metric tensor, $g^\parallel_{\mu\nu}$=${\rm{diag}}(1,0,0,-1)$. 

In contrast to two ordinary (isotropic) viscous coefficients ($\eta$ and $\zeta$ in eq. \eqref{definition}) in the absence of magnetic field, seven viscous coefficients are used to describe the viscous behavior in the presence of a magnetic field. Out of these seven viscous coefficients, five shear viscous coefficients are $\eta$, $\eta_1$, $\eta_2$, $\eta_3$ and $\eta_4$, one bulk or volume viscous coefficient is $\zeta$ and a cross-effect between the ordinary and volume viscosities is $\zeta_1$. So, at finite magnetic field (${\bf B}$ is along a direction, $\bf b=\frac{\bf B}{\rm B}$), the viscous tensor can be written as the linear combination of seven independent tensors \cite{Lifshitz:BOOK'1981}, 
\begin{align}\label{Form1}
\pi_{ij} =& 2 \eta\left(V_{ij}-\frac{1}{3}
\delta_{ij} \nabla \cdot \mathbf V \right) +\zeta 
\delta_{ij} \nabla \cdot \mathbf V\nonumber \\
& +\eta_1\left(2V_{ij}-\delta_{ij}\nabla\cdot\mathbf{V} +
\delta_{ij}V_{kl}b_k b_l-2V_{ik}
b_k b_j-2V_{jk}b_k b_i+b_i b_j\nabla
\cdot\mathbf{V}+b_i b_j V_{kl}b_k b_l\right) 
\nonumber \\
& +2\eta_2\left(V_{ik}b_k b_j+V_{jk}b_k b_i-2b_i b_j V_{kl}b_k b_l
\right)\nonumber \\
& +\eta_3\left(V_{ik}b_{jk}+V_{jk}
b_{ik}-V_{kl}b_{ik}b_j b_l-
V_{kl}b_{jk}b_i b_l\right) \nonumber \\
& +2\eta_4\left(V_{kl}b_{ik}b_j b_l +
V_{kl}b_{jk}b_i b_l\right)\nonumber \\
& +\zeta_1\left(\delta_{ij}V_{kl}b_k b_l+
b_i b_j\nabla\cdot\mathbf{V}\right)
,\end{align}
where $b_{ij}=\epsilon_{ijk}b_k$ and $V_{ij}=\frac{1}{2}
\left(\frac{\partial V_i}{\partial x_j}
+\frac{\partial V_j}{\partial x_i}\right)$. In the above equation, the coefficients of $\eta$, $\eta_1$, $\eta_2$, $\eta_3$ and $\eta_4$ are traceless, whereas the coefficients of $\zeta$ and $\zeta_1$ are having finite trace. The terms containing $\eta$ and $\zeta$ in eq. \eqref{Form1} resemble the terms at $B=0$, thus $\eta$ and $\zeta$ are recognized as the ordinary viscosity coefficients. In case of a plasma, the cross effect between ordinary viscosity and volume viscosity ($\zeta_1$) in eq. \eqref{Form1} vanishes and in the strong magnetic field limit, $\eta_1, \eta_2, \eta_3$ and $\eta_4$ coefficients also vanish, thus converting eq. \eqref{Form1} to a much simpler form, which can be realized through the replacement of the $\eta$-term in the above equation by $\eta_0 \left(3 b_i b_j - \delta_{ij}\right) \left(b_k b_l V_{kl} -\frac{1}{3} \nabla \cdot V\right)$. Thus in Cartesian coordinates, the components of the tensor \eqref{Form1} in a magnetic field (along $z$-direction) are expressed as
\begin{eqnarray}
&&\pi_{xx} = -\eta_0\left(V_{zz}-\frac{1}{3}\nabla\cdot
\mathbf{V}\right)+\eta_1\left(V_{xx}-V_{yy}\right)+2\eta_3 V_{xy}+
\zeta_0\nabla\cdot\mathbf{V} , \\
&&\pi_{yy} = -\eta_0\left(V_{zz}-\frac{1}{3}\nabla\cdot
\mathbf{V}\right)+\eta_1\left(V_{yy}-V_{xx}\right)-2\eta_3V_{xy}+
\zeta_0\nabla\cdot\mathbf{V}, \\
&&\pi_{zz} = 2\eta_0\left(V_{zz}-\frac{1}{3}\nabla\cdot
\mathbf{V}\right)+\zeta_0 \nabla\cdot\mathbf{V}, \\
&&\pi_{xy} = 2\eta_1V_{xy}-\eta_3\left(V_{xx}-V_{yy}\right), \\
&&\pi_{xz} = 2\eta_2 V_{xz}+2\eta_4 V_{yz}, \\
&&\pi_{yz} = 2\eta_2 V_{yz} - 2\eta_4 V_{xz}
.\end{eqnarray}
The presence of strong magnetic field constrains the motion to one spatial dimension (along the direction of magnetic field), which results in the vanishing of the transverse components of the velocity gradient - $V_{xx},V_{yy},V_{xy}$ etc. Thus, the nondiagonal terms of the tensor - $\pi_{xy}$, $\pi_{xz}$ and $\pi_{yz}$ become zero. So the nonvanishing longitudinal components of the viscous tensor are written as
\begin{eqnarray}
&&\pi_{xx} = -\eta_0\left(V_{zz}-\frac{1}{3}\nabla\cdot
\mathbf{V}\right)+\zeta_0\nabla\cdot\mathbf{V} , \\
&&\pi_{yy} = -\eta_0\left(V_{zz}-\frac{1}{3}\nabla\cdot
\mathbf{V}\right)+\zeta_0\nabla\cdot\mathbf{V}, \\
&&\pi_{zz} = 2\eta_0\left(V_{zz}-\frac{1}{3}\nabla\cdot
\mathbf{V}\right)+\zeta_0 \nabla\cdot\mathbf{V}
,\end{eqnarray}
where $\eta_0$ and $\zeta_0$ represent the longitudinal shear and bulk viscosities (with respect to the direction of magnetic field), respectively. The above equations are related to each other and can be grouped in terms of the shear and bulk viscous parts as
\begin{eqnarray}
&&\pi_{zz}=-2\pi_{xx}=-2\pi_{yy}=2\eta_0
\left(V_{zz}-\frac{1}{3} {\nabla\cdot \mathbf{V}|}_z \right), \\ 
&&\pi_{zz}=\pi_{xx}=\pi_{yy}=\zeta_0 {\nabla\cdot\mathbf{V}|}_z
,\end{eqnarray}
respectively. Generalizing the viscous tensor into the relativistic energy-momentum tensor, ${\tilde{T}}^{\mu \nu}$ \cite{Lifshitz:BOOK'1981,Ofengeim:EPL112'2015} in a strong magnetic field, the dissipative part of the relativistic energy-momentum tensor is defined ($\eta_0 \equiv \eta^B$ and $\zeta_0 \equiv \zeta^B$ have been relabelled as artifacts of the strong magnetic field limit) as
\begin{align}\label{FORM (1)}
\nonumber\Delta\tilde{T}^{\mu\nu} =&-\eta^B\left(\frac{\partial{u}^\mu}{\partial\tilde{x}_\nu}+\frac{\partial{u}^\nu}{\partial\tilde{x}_\mu}-{u}^\nu{u}_\lambda\frac{\partial{u}^\mu}{\partial\tilde{x}_\lambda}-{u}^\mu{u}_\lambda\frac{\partial{u}^\nu}{\partial\tilde{x}_\lambda}-\frac{2}{3}\Delta_\parallel^{\mu\nu}\frac{\partial{u}^\lambda}{\partial\tilde{x}^\lambda}\right) \\ & -\zeta^B\Delta_\parallel^{\mu\nu}\frac{\partial{u}^\lambda}{\partial\tilde{x}^\lambda}
~,\end{align}
where $\tilde{x}^\mu=(x^0,0,0,x^3)$, and $\eta^B$ and $\zeta^B$ represent the shear viscosity and the bulk viscosity, respectively in a strong magnetic field. In the local rest frame, the spatial component of velocity is zero, but its spatial derivative remains finite. So, the spatial component of the dissipative part of the relativistic energy-momentum tensor (eq. \eqref{FORM (1)}) in the presence of a strong magnetic field is written \cite{Lifshitz:BOOK'1981,Ofengeim:EPL112'2015} as
\begin{align}\label{definition(eb)}
\nonumber\Delta\tilde{T}^{ij} =& -\eta^B\left(\frac{\partial{u}^i}{\partial\tilde{x}_j}+\frac{\partial{u}^j}{\partial\tilde{x}_i}-\frac{2}{3}\delta^{ij}\frac{\partial{u}^l}{\partial\tilde{x}^l}\right)
-\zeta^B\delta^{ij}\frac{\partial{u}^l}{\partial\tilde{x}^l} \\ =& \nonumber -\eta^B\left(\partial^i{u}^j+\partial^j{u}^i-\frac{2}{3}\delta^{ij}\partial_l{u}^l\right)
-\zeta^B\delta^{ij}\partial_l{u}^l \\ =& -\eta^B{W}^{ij}-\zeta^B\delta^{ij}\partial_l{u}^l
.\end{align}
Comparing equations \eqref{emb3} and \eqref{definition(eb)}, the charged particle (quarks and antiquarks) contribution to the shear viscosity is obtained for a dense QCD medium in the presence of a strong magnetic field as
\be\label{A.S.V.q(eB)}
\eta_q^B &=& \frac{\beta}{8\pi^2}\sum_f g_f ~ |q_fB|\int dp_3~\frac{p_3^4}{\omega_f^2}\left[\tau_f^B ~ f_f^B\left(1-f_f^B\right)+\tau_{\bar{f}}^B ~ \bar{f}_f^B\left(1-\bar{f}_f^B\right)\right]
.\ee
As has been mentioned earlier that gluons are not affected by the presence of magnetic field, so the gluon part of the shear viscosity remains unchanged. Thus, one can add the isotropic gluon part to the anisotropic charged particle part to obtain the total shear viscosity,
\be\label{A.S.V.(eB)}
\nonumber\eta^B &=& \frac{\beta}{8\pi^2}\sum_f g_f ~ |q_fB|\int dp_3~\frac{p_3^4}{\omega_f^2}\left[\tau_f^B ~ f_f^B\left(1-f_f^B\right)+\tau_{\bar{f}}^B ~ \bar{f}_f^B\left(1-\bar{f}_f^B\right)\right] \\ && +\frac{\beta}{30\pi^2} g_g \int d{\rm p}~\frac{{\rm p}^6}{\omega_g^2} ~ \tau_g ~ f_g^{\rm iso}\left(1+f_g^{\rm iso}\right)
.\ee
Similarly, the bulk viscosity due to the charged particle (quarks and antiquarks) contribution is obtained by comparing equations \eqref{emb3} and \eqref{definition(eb)},
\be\label{A.B.V.q(eB)}
\zeta_q^B &=& \sum_f \frac{g_f|q_fB|}{4\pi^2}\int d p_3~\frac{p_3^2}{\omega_f}\left[f_f^B\left(1-f_f^B\right)A_f+\bar{f}_f^B\left(1-\bar{f}_f^B\right)\bar{A}_f\right]
,\ee
where $A_f$ and $\bar{A}_f$ are respectively written as
\begin{eqnarray}
&&A_f=\frac{\tau_f^B}{3T}\left[\frac{p_3^2}{\omega_f}-3\left(\frac{\partial P}{\partial \varepsilon}\right)\omega_f\right] \label{A1i}, \\ 
&&\bar{A}_f=\frac{\tau_{\bar{f}}^B}{3T}\left[\frac{p_3^2}{\omega_f}-3\left(\frac{\partial P}{\partial \varepsilon}\right)\omega_f\right] \label{A1i.1}
.\end{eqnarray}
Implementing the Landau-Lifshitz condition for the calculation of the bulk viscosity and then simplifying, we get
\be\label{A.B.V.q1(eB)}
\zeta_q^B=\frac{\beta}{12\pi^2}\sum_f g_f|q_fB|\int d p_3~\left[\frac{p_3^2}{\omega_f}-3\left(\frac{\partial P}{\partial\varepsilon}\right)\omega_f\right]^2\left[\tau_f^B f^B_f\left(1-f^B_f\right)+\tau^B_{\bar{f}}\bar{f}^B_f\left(1-\bar{f}^B_f\right)\right]
.\ee
The total bulk viscosity is obtained by adding the isotropic gluon part to the anisotropic charged particle part as
\be\label{A.B.V.(eB)}
\nonumber\zeta^B &=& \frac{\beta}{12\pi^2}\sum_f g_f|q_fB|\int d p_3~\left[\frac{p_3^2}{\omega_f}-3\left(\frac{\partial P}{\partial\varepsilon}\right)\omega_f\right]^2\left[\tau_f^B f^B_f\left(1-f^B_f\right)+\tau^B_{\bar{f}}\bar{f}^B_f\left(1-\bar{f}^B_f\right)\right] \\ && +\frac{\beta}{18\pi^2}g_g\int d{\rm p}~{\rm p}^2\left[\frac{{\rm p}^2}{\omega_g}-3\left(\frac{\partial P}{\partial \varepsilon}\right)\omega_g\right]^2 \tau_g ~ f_g^{\rm iso}\left(1+f_g^{\rm iso}\right)
.\ee

\section{Applications}
This section contains some applications of shear and bulk viscosities. We will study the effects of strong magnetic field and finite chemical potential on the interplays between momentum diffusion and thermal diffusion through the Prandtl number, between momentum diffusion and characteristic length scale of the system through the Reynolds number, and between shear viscosity and bulk viscosity through the ratio $\zeta/\eta$ in subsections 3.1, 3.2 and 3.3, respectively. 

\subsection{Prandtl number}
The relative importance between the momentum diffusion and the thermal diffusion in a medium can be explained through the Prandtl number (Pl), 
\begin{equation}\label{Pl}
{\rm Pl}=\frac{\eta/\rho}{\kappa/C_p}
~,\end{equation}
where $\rho$, $\kappa$ and $C_p$ represent the mass density, the thermal conductivity and the specific heat at constant pressure, respectively. The Prandtl number is helpful in understanding the effects of thermal conductivity and shear viscosity on the sound attenuation in a medium. For various systems different Prandtl numbers have been reported, such as, for strongly coupled liquid helium, Pl is approximately 2.5 \cite{T:RPP72'2009}, for a nonrelativistic conformal holographic fluid, Pl is 1.0 \cite{T:RPP72'2009,Rangamani:JHEP01'2009} and for a dilute atomic Fermi gas at high temperature, Pl is $\frac{2}{3}$ \cite{Braby:PRA82'2010}. Through the Prandtl number one
can get the knowledge about the sound attenuation in a medium that mainly describes the energy loss due to the sound propagation in that medium. Small Pl means that the sound attenuation is mainly dominated by the thermal diffusion while for large Pl, the sound attenuation is prominently governed by the momentum diffusion. The value 1 of the Prandtl number signifies equal contributions of both the diffusions on the sound attenuation. In this work, we want to observe how the presence of strong magnetic field and finite chemical potential affects the Prandtl number for a hot QCD matter. 

In order to calculate the Prandtl number for the hot QCD matter in the presence of both strong magnetic field and chemical potential, we require the expressions of shear viscosity, thermal conductivity, mass density and specific heat at constant pressure. We have recently calculated the thermal conductivity in the similar environment \cite{Rath:EPJC80'2020}, so, we have taken the results on $\kappa$ from our work (appendix \ref{A.T.C.}). Then we have determined the mass density from the product of the number densities of quarks, antiquarks and gluons with their respective quasiparticle masses as
\begin{equation}\label{M.D.}
\rho=\sum_f m_f\left(n_f+\bar{n}_f\right)+m_gn_g
~.\end{equation}
Thus, the expressions of $\rho$ for an isotropic dense QCD medium and for a dense QCD medium in the presence of a strong magnetic field are written as
\begin{eqnarray}
\rho^{\rm iso} &=& \frac{1}{2\pi^2}\sum_f m_f g_f \int d{\rm p}~{\rm p}^2\left[f_f^{\rm iso}+\bar{f}_f^{\rm iso}\right]+\frac{1}{2\pi^2}m_g g_g\int d{\rm p}~{\rm p}^2f_g^{\rm iso}, \label{iso.(M.D.)} \\ 
\rho^B &=& \frac{1}{4\pi^2}\sum_f m_f g_f|q_fB|\int d p_3\left[f_f^B+\bar{f}_f^B\right]+\frac{1}{2\pi^2}m_g g_g\int d{\rm p}~{\rm p}^2f_g^{\rm iso} \label{baniso.(M.D.)}
,\end{eqnarray}
respectively. Next, we have calculated $C_p$ from the energy density and the pressure using the following thermodynamic relation, 
\begin{equation}\label{cp}
C_p=\frac{\partial(\varepsilon+P)}{\partial T}
~,\end{equation}
where $\varepsilon$ and $P$ are determined for the dense QCD medium in the absence and presence of strong magnetic field in appendix \ref{A.(e,p)} using the kinetic theory. So, the expressions of $C_p$ for an isotropic dense QCD medium and for a dense QCD medium in the presence of a strong magnetic field are obtained as
\begin{eqnarray}
\nonumber C_p^{\rm iso} &=& \frac{\beta^2}{6\pi^2}\sum_f g_f \int d{\rm p}~{\rm p}^2\left(\frac{{\rm p}^2}{\omega_f}+3\omega_f\right)\left[\left(\omega_f-\mu\right)f_f^{\rm iso}\left(1-f_f^{\rm iso}\right)+\left(\omega_f+\mu\right)\bar{f}_f^{\rm iso}\left(1-\bar{f}_f^{\rm iso}\right)\right] \\ && +\frac{\beta^2}{6\pi^2}g_g\int d{\rm p}~{\rm p}^2\left({\rm p}^2+3\omega_g^2\right)f_g^{\rm iso}\left(1+f_g^{\rm iso}\right), \label{iso.cp} \\  
\nonumber C_p^B &=& \frac{\beta^2}{4\pi^2}\sum_f g_f|q_fB|\int d p_3\left(\frac{p_3^2}{\omega_f}+\omega_f\right)\left[\left(\omega_f-\mu\right)f_f^B\left(1-f_f^B\right)+\left(\omega_f+\mu\right)\bar{f}_f^B\left(1-\bar{f}_f^B\right)\right] \\ && +\frac{\beta^2}{6\pi^2}g_g\int d{\rm p}~{\rm p}^2\left({\rm p}^2+3\omega_g^2\right)f_g^{\rm iso}\left(1+f_g^{\rm iso}\right) \label{baniso.cp}
,\end{eqnarray}
respectively. Substituting the values of $\eta$, $\kappa$, $\rho$ and $C_p$ in eq. \eqref{Pl}, Pl is calculated. Our observation on the Prandtl number for a hot QCD matter in the presence of both strong magnetic field and finite chemical potential with the quasiparticle model has been described in subsection 5.2. 

\subsection{Reynolds number}
The Reynolds number (Rl) is an important quantity, that helps in assessing the magnitude of the kinematic viscosity (${\eta}/{\rho}$) of a liquid and is defined by
\begin{equation}\label{Rl}
{\rm Rl}=\frac{Lv}{\eta/\rho}
~,\end{equation}
where $L$ and $v$ denote the characteristic length and the velocity of the flow of liquid, respectively. As the Reynolds number is the ratio of the product of characteristic length and velocity ($Lv$) to the kinematic viscosity of a system, its magnitude gives the information about the fluidity of that system. In hydrodynamics, Rl describes the motion of the fluid and indicates when the laminar flow gets converted into the turbulent flow. This conversion to the turbulent flow happens when ${\rm Rl}\gg1$ ({\em i.e.} in the thousands) or when the kinematic viscosity is very small as compared to the product of characteristic length and velocity of the system. Different systems have reported different Reynolds numbers, such as, Rl of quark matter is estimated to be about 10 using the Kubo formula and NJL model \cite{Fukutome:PTP119'2008}, the (3+1)-dimensional fluid dynamical model has reported the value of Rl in the range 3-10 for QGP \cite{Csernai:PRC85'2012} and the holographic model has estimated the upper bound of Rl as approximately 20 \cite{McInnes:NPB921'2017}. The small value of Rl describes the QGP as a viscous system and the nature of its flow remains laminar. Thus, Rl gives the information about the magnitude of the viscosity and the fluid dynamic characteristics of a system. In our work, we intend to analyse the effects of strong magnetic field and finite chemical potential on the Reynolds number for a hot QCD matter. In the calculation, we have set $v\simeq 1$ and $L=4$ fm. Our observation on the Reynolds number for a hot QCD matter in the presence of both strong magnetic field and finite chemical potential with the quasiparticle model has been described in subsection 5.3. 

\subsection{Relative behavior between shear viscosity and bulk viscosity}
The competition between the shear viscous effect and the bulk viscous effect in a system can be understood by studying the ratio $\zeta/\eta$ in that system. Previously some observations have found different behaviors of $\zeta/\eta$ with temperature depending upon the system dynamics at different coupling regimes. For an interacting scalar field, $\zeta/\eta$ was calculated to be $15\left(\frac{1}{3}-c_s^2\right)^2$ ($c_s^2$ is the square of the speed of sound) \cite{Horsley:NPB280'1987}. The study at perturbative limit also shows that $\zeta/\eta$ is nearly equal to $15\left(\frac{1}{3}-c_s^2\right)^2$ for a hot QCD medium \cite{Arnold:PRD74'2006}, whereas that for a strongly coupled gauge theory plasma is nearly equal to $2\left(\frac{1}{3}-c_s^2\right)$ \cite{Buchel:PLB663'2008}. Thus, going from the perturbative regime to the nonperturbative regime, the ratio $\zeta/\eta$ undergoes a gradual change. In the holographic model \cite{Buchel:NPB820'2009}, for QGP in the high temperature regime, $\frac{\zeta}{\eta}<\frac{1}{2}$ and in the deconfinement transition region, $\zeta/\eta$ approaches 0.6. Similar behavior of $\zeta/\eta$ has also been observed in ref. \cite{U:JHEP0912'2009} using the holographic model, where $\zeta$ remains lower than $\eta$ for high temperatures and gets risen near the deconfinement transition region. In case of a quasigluon plasma \cite{Bluhm:PLB709'2012}, for $T\geq1.5T_c$ ($T_c$ is the critical temperature of phase transition), the ratio $\zeta/\eta$ shows the perturbative QCD-like behavior ($\frac{\zeta}{\eta}=15\left(\frac{1}{3}-c_s^2\right)^2$), whereas at $T\simeq T_c$, the nonperturbative effect comes into picture, so in particular, for $T\geq1.02T_c$, $\eta$ is larger than $\zeta$ and for $T\rightarrow1.02T_c$, the ratio $\zeta/\eta$ approaches 0.78. According to the observations in the hadron gas \cite{Prakash:PR227'1993,Davesne:PRC53'1996,Chen:PRC79'2009}, the bulk viscosity also remains much smaller than the shear viscosity. In the Chapman-Enskog method \cite{Mitra:PRD96'2017}, $\zeta/\eta$ for a hot QCD medium was observed as a function of temperature using two different scalings, where with scaling $15\left(\frac{1}{3}-c_s^2\right)^2$, the ratio $\zeta/\eta$ increases with the temperature at high temperature regime, whereas that with scaling $2\left(\frac{1}{3}-c_s^2\right)$ was found to decrease with the temperature. Our observation on the ratio $\zeta/\eta$ for a hot QCD matter in the presence of both strong magnetic field and finite chemical potential with the quasiparticle model has been described in subsection 5.4. 

\section{Quasiparticle model for hot and dense QCD medium in the presence of a strong magnetic field}
The interaction of a particle with other particles in a thermal medium results in the emergence of its quasiparticle mass. For a hot and dense QCD medium, this thermally generated mass or quasiparticle mass depends on temperature and chemical potential, and in the additional presence of strong magnetic field, this mass depends on temperature, chemical potential as well as magnetic field. So, in the quasiparticle model (QPM), the QGP consists of massive noninteracting quasiparticles. The quasiparticle mass can be derived using different models, {\em viz.}, the Nambu-Jona-Lasinio (NJL) and Polyakov NJL based quasiparticle models \cite{Fukushima:PLB591'2004,Ghosh:PRD73'2006,Abuki:PLB676'2009}, quasiparticle model with Gribov-Zwanziger quantization \cite{Su:PRL114'2015,Florkowski:PRC94'2016}, thermodynamically consistent quasiparticle model \cite{Bannur:JHEP0709'2007} etc. In a thermal medium at finite but small chemical potential ($\mu_f$) and zero magnetic field, the thermal mass (squared) of quark of $f$th flavor is given \cite{Braaten:PRD45'1992,Peshier:PRD66'2002} by
\be\label{Q.P.M.}
m_{fT}^2=\frac{g^{\prime2}T^2}{6}\left(1+\frac{\mu_f^2}{\pi^2T^2}\right)
,\ee
where $g^\prime$ represents the running coupling at finite temperature and finite chemical potential. In the semiclassical transport theory, the thermal mass (squared) of gluon is defined \cite{Kelly:PRD50'1994,Litim:PR364'2002} as
\be\label{Q.P.M.(definition of gluon mass)}
\nonumber m_{gT}^2 &=& -g^{\prime2}N_c\int\frac{d^3{\rm k}}{(2\pi)^3}\frac{\partial f_g^{\rm iso}}{\partial{\rm k}}-\frac{g^{\prime2}N_f}{2}\int\frac{d^3{\rm k}}{(2\pi)^3}\left(\frac{\partial f_f^{\rm iso}}{\partial{\rm k}}+\frac{\partial \bar{f}_f^{\rm iso}}{\partial{\rm k}}\right) \\ &=& \nonumber\frac{g^{\prime2}N_c}{2\pi^2T}\int d{\rm k}~\frac{{\rm k}^3}{\omega_g}f_g^{\rm iso}\left(1+f_g^{\rm iso}\right)+\frac{g^{\prime2}N_f}{4\pi^2T}\int d{\rm k}~\frac{{\rm k}^3}{\omega_f}\left[f_f^{\rm iso}\left(1-f_f^{\rm iso}\right)\right. \\ && \left.+\bar{f}_f^{\rm iso}\left(1-\bar{f}_f^{\rm iso}\right)\right]
,\ee
which can be calculated in the hard thermal loop approximation and thus, the simplified form of $m_{gT}^2$ for small chemical potential is written \cite{Peshier:PRD66'2002,Blaizot:PRD72'2005,Berrehrah:PRC89'2014} as
\be\label{Q.P.M.(Gluon mass)}
m_{gT}^2=\frac{g^{\prime2}T^2}{6}\left(N_c+\frac{N_f}{2}+\frac{3}{2\pi^2T^2}\sum_f\mu_f^2\right)
.\ee

However, in a strong magnetic field, eq. \eqref{Q.P.M.(definition of gluon mass)} gets modified into
\be\label{Q.P.M.eB(definition of gluon mass)}
m_{gT,B}^2 &=& -g^{\prime2}N_c\int\frac{d^3{\rm k}}{(2\pi)^3}\frac{\partial f_g^{\rm iso}}{\partial{\rm k}}-\frac{g^2}{8\pi^2}\sum_f |q_fB|\int dk_z\left(\frac{\partial f_f^B}{\partial k_z}+\frac{\partial \bar{f}_f^B}{\partial k_z}\right)
.\ee
After solving the above equation, we get the thermal gluon mass (squared) in the presence of both strong magnetic field and finite chemical potential as
\be\label{Q.P.M.eB(Gluon mass)}
m_{gT,B}^2 &=& \frac{g^{\prime2}T^2N_c}{6}+\frac{g^2}{8\pi^2T}\sum_f|q_fB|\int dk_z~\frac{k_z}{\omega_f}\left[f_f^B\left(1-f_f^B\right)+\bar{f}_f^B\left(1-\bar{f}_f^B\right)\right]
.\ee
Similarly, the strong magnetic field also modifies the thermal quark mass (eq. \eqref{Q.P.M.}), which can be calculated by taking $p_0=0, p_z\rightarrow 0$ limit of the effective quark propagator. For this purpose, one needs to first determine the effective quark propagator from the self-consistent Schwinger-Dyson equation in an ambience of strong magnetic field, 
\be\label{S.D.E.}
S^{-1}(p_\parallel)=\gamma^\mu p_{\parallel\mu}-\Sigma(p_\parallel)
~.\ee
The quark self-energy in a strong magnetic field is written as
\begin{eqnarray}\label{Q.S.E.}
\Sigma(p)=-\frac{4}{3} g^{2}i\int{\frac{d^4k}{(2\pi)^4}}\left[\gamma_\mu {S(k)}\gamma^\mu{D(p-k)}\right]
,\end{eqnarray}
where $g$ denotes the running coupling in the strong magnetic field regime \cite{Andreichikov:PRL110'2013,Ferrer:PRD91'2015,Ayala:PRD98'2018,Viscosities}. In this regime, the quark propagator $S(k)$ in vacuum is given \cite{Schwinger:PR82'1951,Tsai:PRD10'1974} by
\be\label{q. propagator}
S(k)=ie^{-\frac{k^2_\perp}{|q_fB|}}\frac{\left(\gamma^0 k_0-\gamma^3 k_z+m_f\right)}{k^2_\parallel-m^2_f}\left(1-\gamma^0\gamma^3\gamma^5\right)
,\ee
where the metric tensors and the four-vectors are defined as
\begin{eqnarray*}
&& g^{\mu\nu}_\perp={\rm{diag}}(0,-1,-1,0), ~~ g^{\mu\nu}_\parallel={\rm{diag}}(1,0,0,-1), \\ 
&& k_{\perp\mu}\equiv(0,k_x,k_y,0), ~~ k_{\parallel\mu}\equiv(k_0,0,0,k_z)
.\end{eqnarray*}
The propagator of electrically neutral gluon in vacuum remains unaffected by the magnetic field and carries the following form as in the absence of magnetic field, 
\be
\label{g. propagator}
D^{\mu \nu} (p-k)=\frac{ig^{\mu \nu}}{(p-k)^2}
~.\ee
Substituting quark \eqref{q. propagator} and gluon \eqref{g. propagator} propagators in eq. \eqref{Q.S.E.}, we have calculated the quark self-energy in the imaginary time formalism at strong magnetic field and finite chemical potential, and its approximated form for $T>\mu_f$ is written \cite{Rath:EPJC80'2020} as
\begin{eqnarray}\label{Q.S.E.(3)}
\nonumber\Sigma(p_\parallel) &\approx& \frac{g^2|q_fB|}{3\pi^2}\left[\frac{\pi T}{2m_f}-\ln(2)+\frac{7\mu_f^2\zeta(3)}{8\pi^2T^2}-\frac{31\mu_f^4\zeta(5)}{32\pi^4T^4}\right] \\ && \times\left[\frac{\gamma^0p_0}{p_\parallel^2}+\frac{\gamma^3p_z}{p_\parallel^2}+\frac{\gamma^0\gamma^5p_z}{p_\parallel^2}+\frac{\gamma^3\gamma^5p_0}{p_\parallel^2}\right]
,\end{eqnarray}
where $\zeta(s)$ is the Riemann zeta function with $s=3,5$ here. The covariant structure of the quark self-energy for a thermal medium in the presence of a magnetic field is written \cite{Karmakar:PRD99'2019,Viscosities} as
\begin{equation}\label{general q.s.e.(11)}
\Sigma(p_\parallel)=A\gamma^\mu u_\mu+B\gamma^\mu b_\mu+C\gamma^5\gamma^\mu u_\mu+D\gamma^5\gamma^\mu b_\mu
~,\end{equation}
where $u^\mu$ (1,0,0,0) and $b^\mu$ (0,0,0,-1) are the directions of heat bath and magnetic field, respectively, and the form factors $A$, $B$, $C$ and $D$ are determined in the strong magnetic field limit as
\begin{eqnarray}
&&A=\frac{g^2|q_fB|}{3\pi^2}\left[\frac{\pi T}{2m_f}-\ln(2)+\frac{7\mu_f^2\zeta(3)}{8\pi^2T^2}-\frac{31\mu_f^4\zeta(5)}{32\pi^4T^4}\right]\frac{p_0}{p_\parallel^2} ~, \\ 
&&B=\frac{g^2|q_fB|}{3\pi^2}\left[\frac{\pi T}{2m_f}-\ln(2)+\frac{7\mu_f^2\zeta(3)}{8\pi^2T^2}-\frac{31\mu_f^4\zeta(5)}{32\pi^4T^4}\right]\frac{p_z}{p_\parallel^2} ~, \\ 
&&C=-\frac{g^2|q_fB|}{3\pi^2}\left[\frac{\pi T}{2m_f}-\ln(2)+\frac{7\mu_f^2\zeta(3)}{8\pi^2T^2}-\frac{31\mu_f^4\zeta(5)}{32\pi^4T^4}\right]\frac{p_z}{p_\parallel^2} ~, \\ 
&&D=-\frac{g^2|q_fB|}{3\pi^2}\left[\frac{\pi T}{2m_f}-\ln(2)+\frac{7\mu_f^2\zeta(3)}{8\pi^2T^2}-\frac{31\mu_f^4\zeta(5)}{32\pi^4T^4}\right]\frac{p_0}{p_\parallel^2}
~.\end{eqnarray}
With $C=-B$ and $D=-A$, eq. \eqref{general q.s.e.(11)} can be written in terms of the chiral projection operators as
\begin{equation}
\Sigma(p_\parallel)=P_R\left[(A-B)\gamma^\mu u_\mu+(B-A)\gamma^\mu b_\mu
\right]P_L+P_L\left[(A+B)\gamma^\mu u_\mu+(B+A)\gamma^\mu b_\mu\right]P_R
~.\end{equation}
Finally, from the Schwinger-Dyson equation \eqref{S.D.E.}, the effective mass (squared) of quark in the presence of strong magnetic field, finite temperature and finite chemical potential is obtained (in appendix \ref{Thermal mass (eB, T, C.P.)}) by taking the $p_0=0, p_z\rightarrow 0$ limit \cite{Rath:EPJC80'2020} as
\begin{eqnarray}\label{Mass}
m_{fT,B}^2=\frac{g^2|q_fB|}{3\pi^2}\left[\frac{\pi T}{2m_f}-\ln(2)+\frac{7\mu_f^2\zeta(3)}{8\pi^2T^2}-\frac{31\mu_f^4\zeta(5)}{32\pi^4T^4}\right]
,\end{eqnarray}
which depends on temperature, chemical potential and magnetic field. 

Our calculations are done in the strong magnetic field limit, where the energy scale related to the magnetic field dominates over the energy scales related to the temperature and chemical potential. So the magnetic field is set at $15$ $m_{\pi}^2$, the temperature is taken in the range $0.16$ GeV - $0.4$ GeV to satisfy the condition $eB \gg T^2$, and the chemical potential is fixed at $0.06$ GeV to satisfy the conditions $eB \gg \mu^2$ and $T^2 \gg \mu^2$. We also note that, all flavors are assigned the same chemical potential ($\mu_f=\mu$). In section 5, we will discuss the results in the quasiparticle model, where the temperature and chemical potential-dependent quark \eqref{Q.P.M.} and gluon \eqref{Q.P.M.(Gluon mass)} masses are used for the dense QCD medium in the absence of magnetic field, and the temperature, chemical potential and magnetic field-dependent quark \eqref{Mass} and gluon \eqref{Q.P.M.eB(Gluon mass)} masses are used for the dense QCD medium in the presence of a strong magnetic field. 

\begin{figure}[]
\begin{center}
\begin{tabular}{c c}
\includegraphics[width=6.3cm]{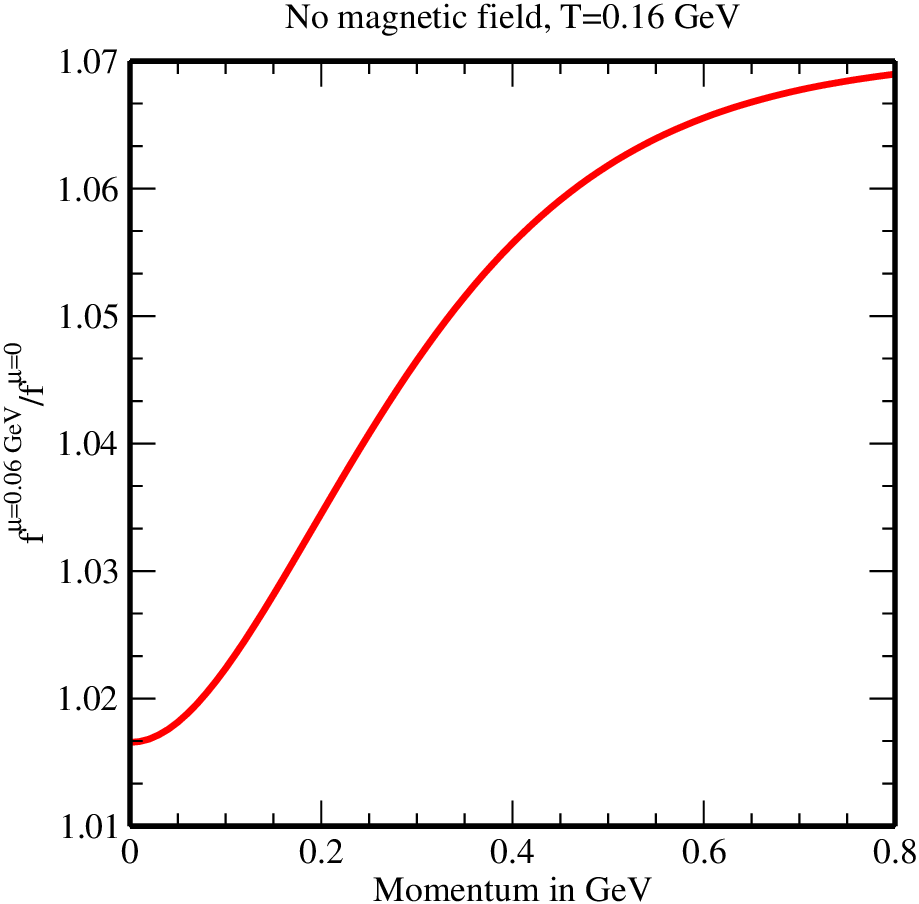}&
\hspace{0.423 cm}
\includegraphics[width=6.3cm]{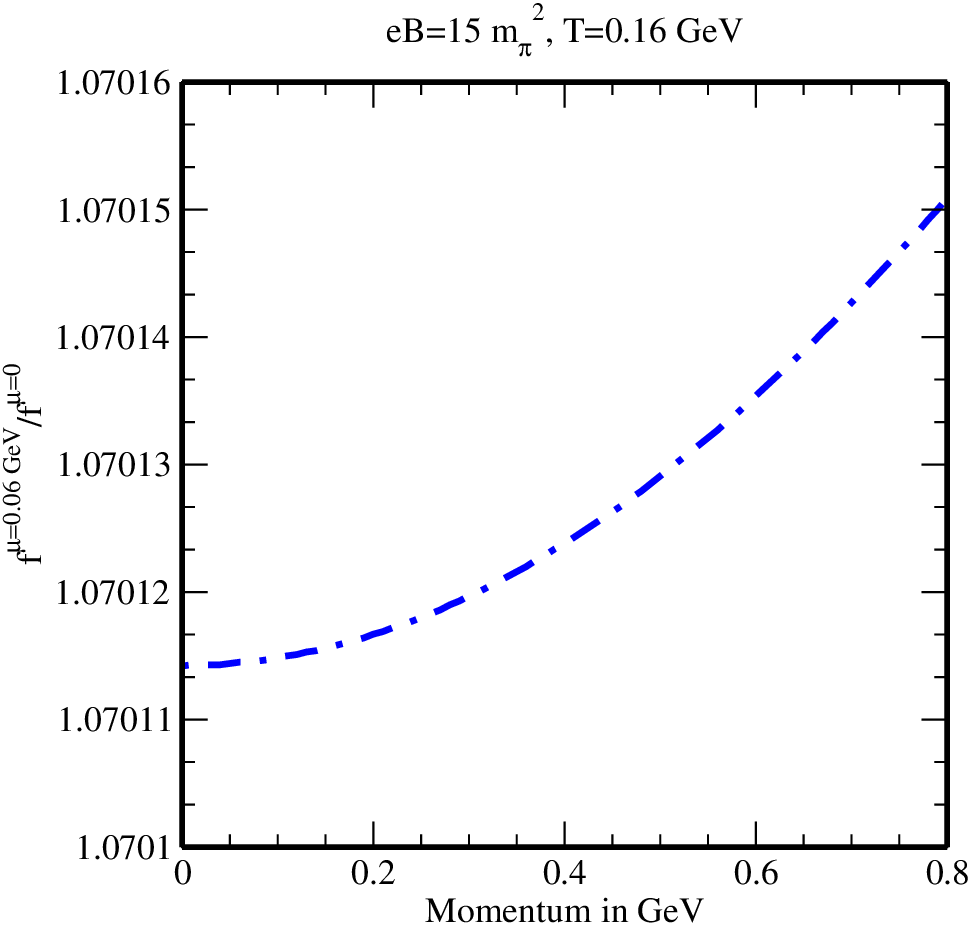} \\
a & b 
\end{tabular}
\caption{Variations of the ratio of quark distribution 
function at $\mu=0.06$ GeV to its value at $\mu=0$ 
with momentum at low temperature in the (a) absence 
and (b) presence of strong magnetic field.}\label{dfup.1}
\end{center}
\end{figure}

\begin{figure}[]
\begin{center}
\begin{tabular}{c c}
\includegraphics[width=6.3cm]{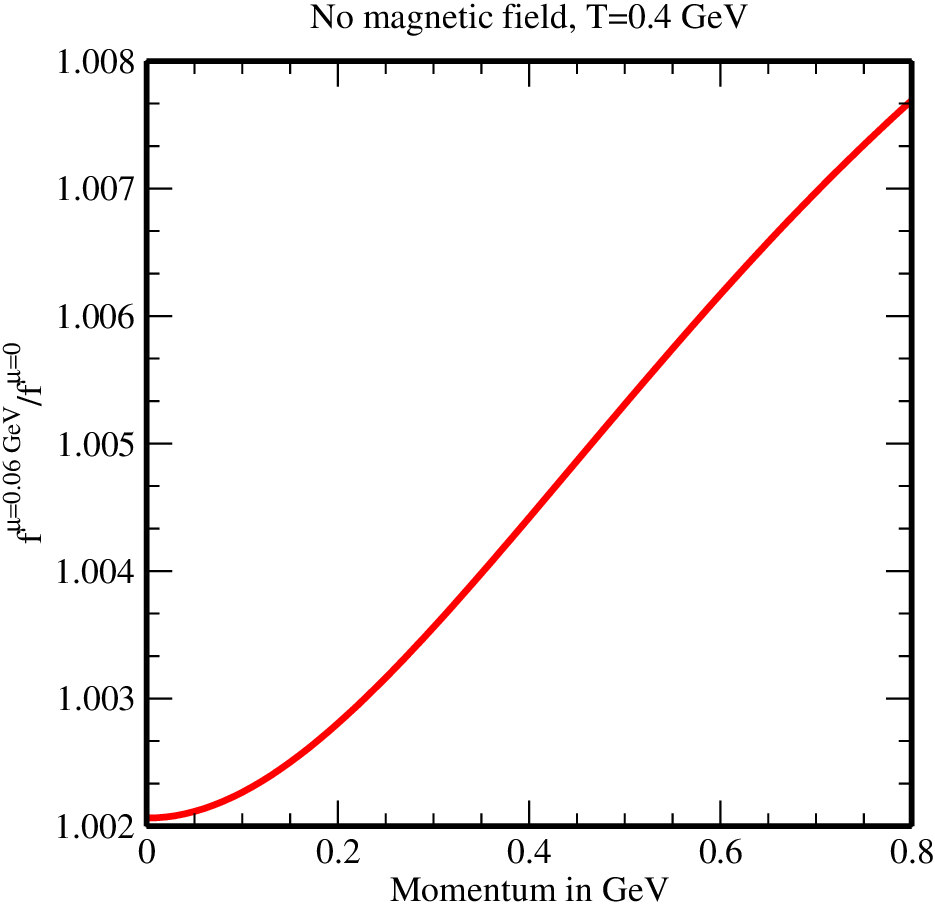}&
\hspace{0.423 cm}
\includegraphics[width=6.3cm]{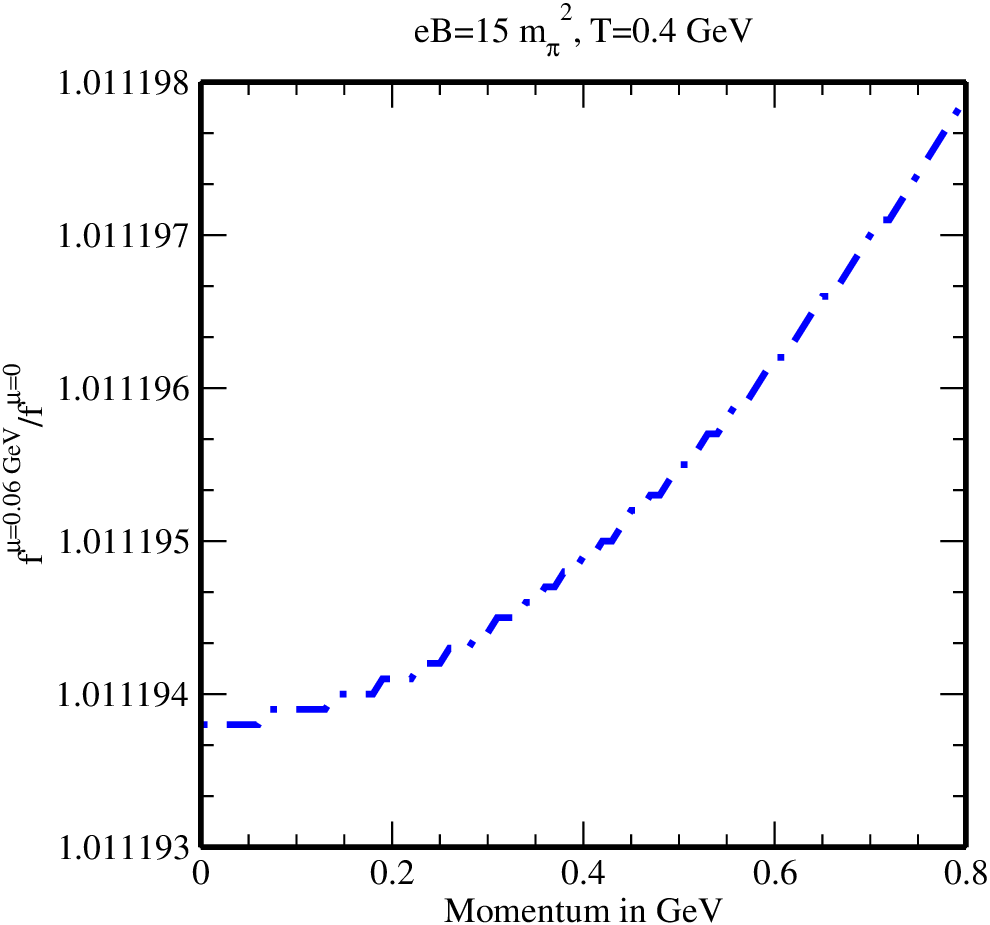} \\
a & b 
\end{tabular}
\caption{Variations of the ratio of quark distribution 
function at $\mu=0.06$ GeV to its value at $\mu=0$ 
with momentum at high temperature in the (a) absence 
and (b) presence of strong magnetic field.}\label{dfup.2}
\end{center}
\end{figure}

In kinetic theory, the distribution function plays an important role in 
understanding the properties of various transport coefficients. So, 
before discussing the results on viscous properties of the hot QCD 
matter in the presence of both strong magnetic field and finite chemical 
potential, the knowledge on the distribution functions of particles 
in the similar environment needs to be acquired. For this purpose, we 
have explored the distribution function for $u$ quark at $\mu=0.06$ GeV 
in unit of its value at $\mu=0$ within the quasiparticle model as a 
function of momentum in low temperature (figure \ref{dfup.1}) and 
high temperature (figure \ref{dfup.2}) scenarios. In all scenarios, 
ratio $f^{\mu=0.06 ~ {\rm GeV}}/f^{\mu=0}$ is slightly 
greater than 1, therefore, the distribution function at finite chemical 
potential is larger than its counterpart at zero chemical potential. We 
have also noticed that, in the absence as well as in the presence of strong 
magnetic field, although the ratio is greater than 1, but it gets 
decreased at high temperature (figures  \ref{dfup.2}a and \ref{dfup.2}b) as 
compared to the low temperature case 
(figures \ref{dfup.1}a and \ref{dfup.1}b). Thus, with the rise 
of temperature, the distribution function becomes increased. However, the 
chemical potential has marginal effect on the distribution function at 
higher temperatures. 

\section{Results and discussions}
This section is devoted to the discussions on the results regarding 
shear and bulk viscosities, Prandtl number, Reynolds number and 
relative behavior between shear viscosity and bulk viscosity through the 
ratio of bulk viscosity to shear viscosity. 

\subsection{Shear and bulk viscosities}
\begin{figure}[]
\begin{center}
\begin{tabular}{c c}
\includegraphics[width=6.3cm]{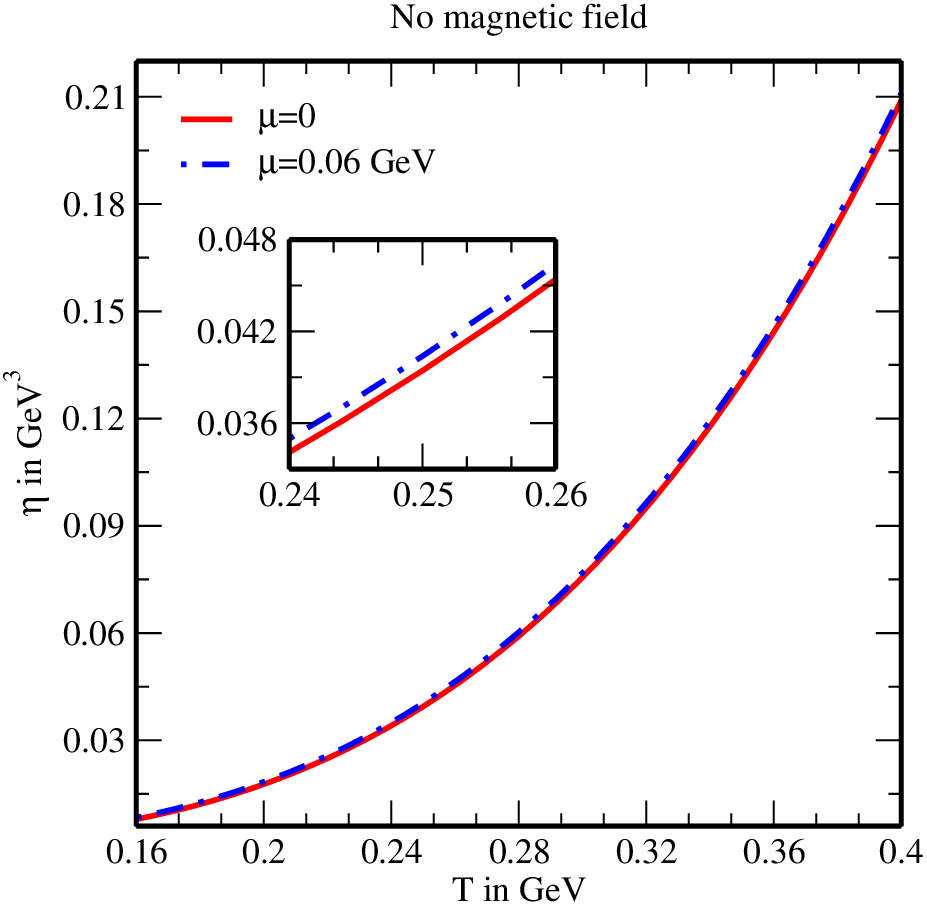}&
\hspace{0.74 cm}
\includegraphics[width=6.3cm]{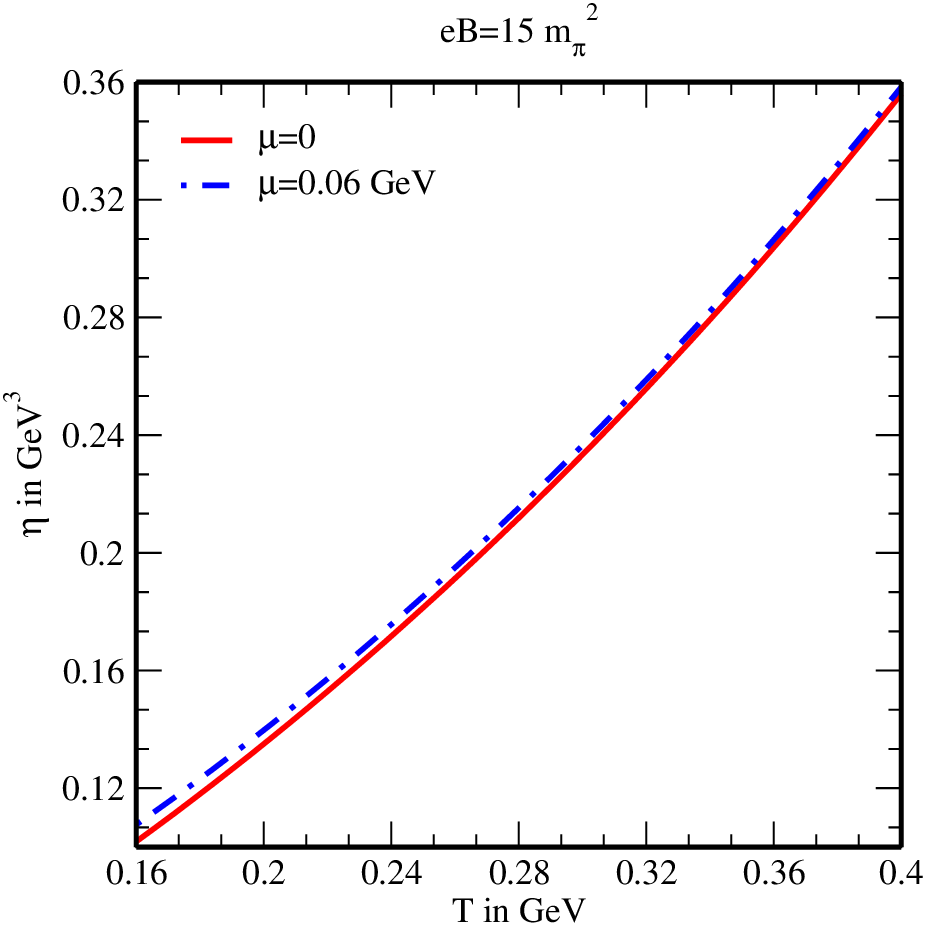} \\
a & b
\end{tabular}
\caption{Variations of the shear viscosity with temperature 
in the (a) absence and (b) presence of strong magnetic field.}\label{Fig.1}
\end{center}
\end{figure}

\begin{figure}[]
\begin{center}
\begin{tabular}{c c}
\includegraphics[width=6.3cm]{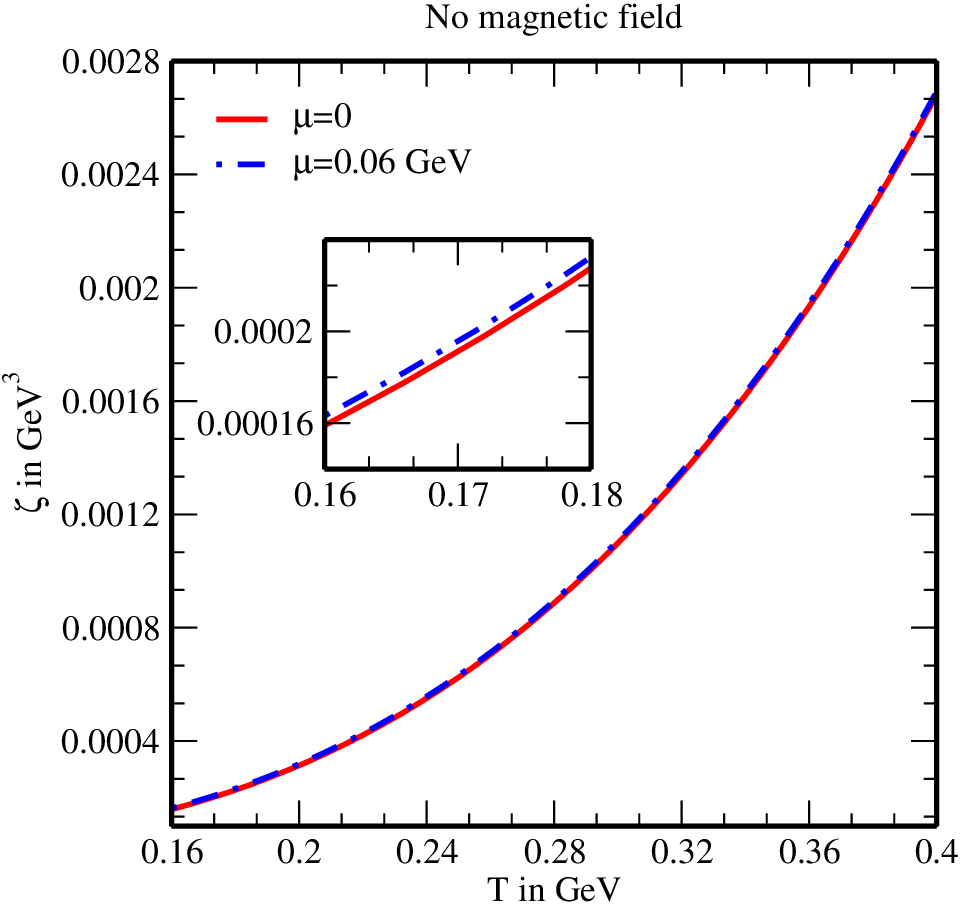}&
\hspace{0.74 cm}
\includegraphics[width=6.3cm]{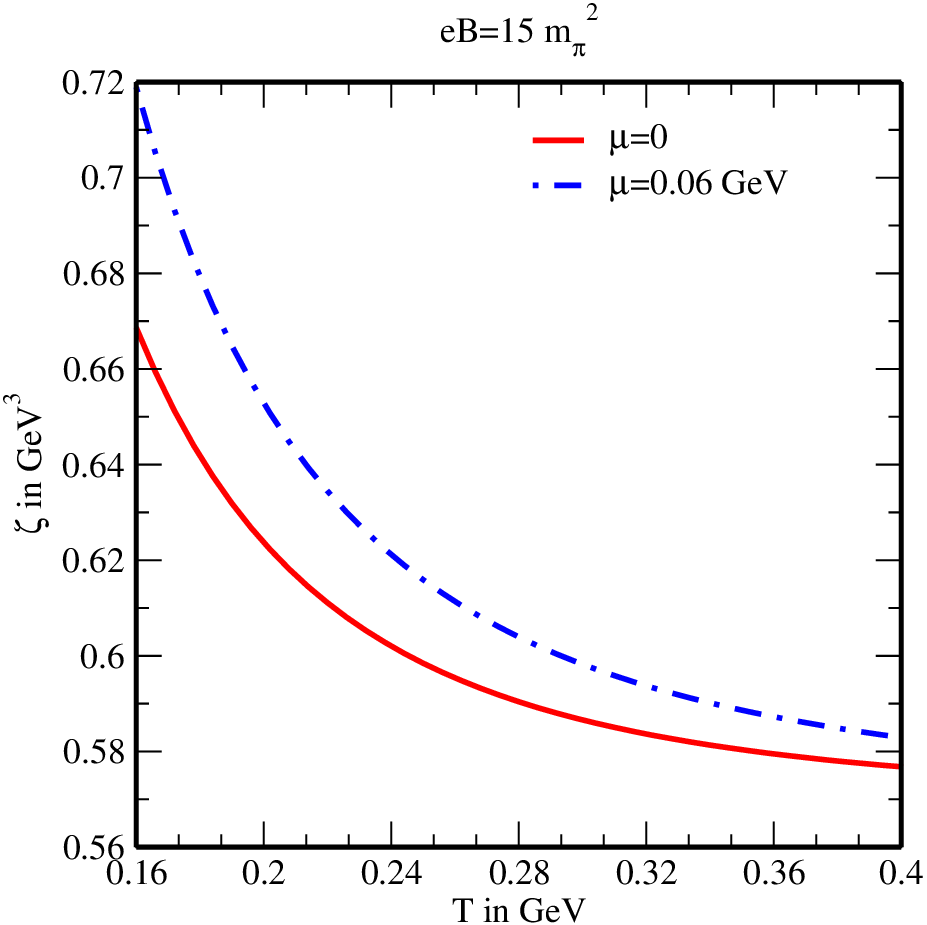} \\
a & b
\end{tabular}
\caption{Variations of the bulk viscosity with temperature 
in the (a) absence and (b) presence of strong magnetic field.}\label{Fig.2}
\end{center}
\end{figure}

From figures \ref{Fig.1} and \ref{Fig.2}, it is seen that both 
shear viscosity and bulk viscosity become increased in the presence 
of a strong magnetic field as compared to their isotropic 
counterparts in the absence of magnetic field. There are two main 
reasons behind the larger magnitudes of $\eta$ and $\zeta$: the 
distribution function and the dispersion relation. The emergence 
of strong magnetic field reduces the dynamics of charged particles 
from three spatial dimensions to one spatial dimension and 
squeezes the phase space. Therefore, the distribution 
function gets stretched along the direction 
longitudinal to the magnetic field and the dispersion 
relation becomes modified. Further, the phase space integral is 
bifurcated into longitudinal and transverse parts with 
respect to the direction of magnetic field, where a 
factor $|q_fB|$ is obtained from the transverse part. 
In addition, in the quasiparticle model, the effects 
of magnetic field, temperature and chemical potential 
are incorporated in the distribution functions through the 
effective masses of particles. Moreover, the partial 
derivative of pressure with respect to energy density in the 
bulk viscosity expression contains the magnetic field 
dependence, unlike that in the absence of magnetic field. As 
a result, $\eta$ and $\zeta$ calculated in an ambience of strong 
magnetic field depend explicitly on magnetic field, chemical 
potential and temperature, whereas in the absence of magnetic 
field, they depend on temperature and chemical potential. Therefore, 
in the strong magnetic field limit ($|q_fB| \gg T^2$), where the 
energy scale related to the magnetic field prevails over the energy 
scales related to the temperature and the chemical potential, $\eta$ and $\zeta$ are 
more sensitive to the magnetic field and as a result, the shear and bulk viscosities 
get enhanced. 

However, with the increase of temperature, shear viscosity increases nearly 
linearly (figure \ref{Fig.1}b) as compared to the slightly nonlinear 
increase in the medium with no magnetic field (figure \ref{Fig.1}a), 
whereas bulk viscosity decreases (figure \ref{Fig.2}b), contrary to its 
increase in the absence of magnetic field (figure \ref{Fig.2}a). This 
behavior of $\zeta$ can be understood from the fact that, in the strong 
magnetic field limit, temperature acts as the weak energy scale, thus it 
may leave a reverse effect on $\zeta$, whereas in the absence of magnetic 
field, temperature acts as the dominant energy scale, so its influence 
on $\zeta$ is more conspicuous in comparison to that in the strong 
magnetic field regime. 

The presence of finite chemical potential also increases the values 
of shear and bulk viscosities for the medium with no magnetic field 
as well as for the medium under the influence of a strong magnetic 
field. The effect of chemical potential on $\eta$ and $\zeta$ is 
more evident at strong magnetic field (figures \ref{Fig.1}b and 
\ref{Fig.2}b) than at zero magnetic field (figures \ref{Fig.1}a and 
\ref{Fig.2}a). Thus, both strong magnetic field and finite chemical 
potential facilitate the transports of momentum across and along the 
layer. From figures \ref{Fig.1} and \ref{Fig.2}, we have also found 
that, the deviations of shear and bulk viscosities at finite 
chemical potential from the counterparts at zero chemical potential 
are more perceptible at low temperature than those at high temperature, 
which can be comprehended from the observations on the distribution 
functions at low and high temperatures (figures \ref{dfup.1} and \ref{dfup.2}, respectively). 

\subsection{Prandtl number}
\begin{figure}[]
\begin{center}
\begin{tabular}{c c}
\includegraphics[width=6.3cm]{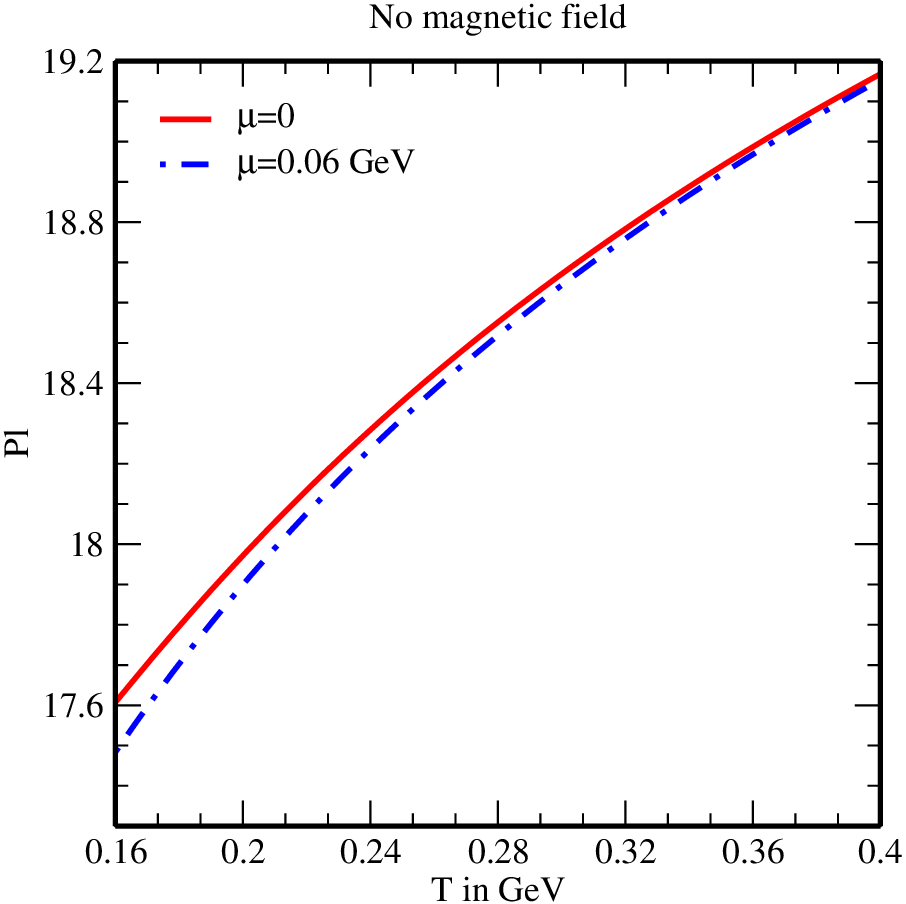}&
\hspace{0.74 cm}
\includegraphics[width=6.3cm]{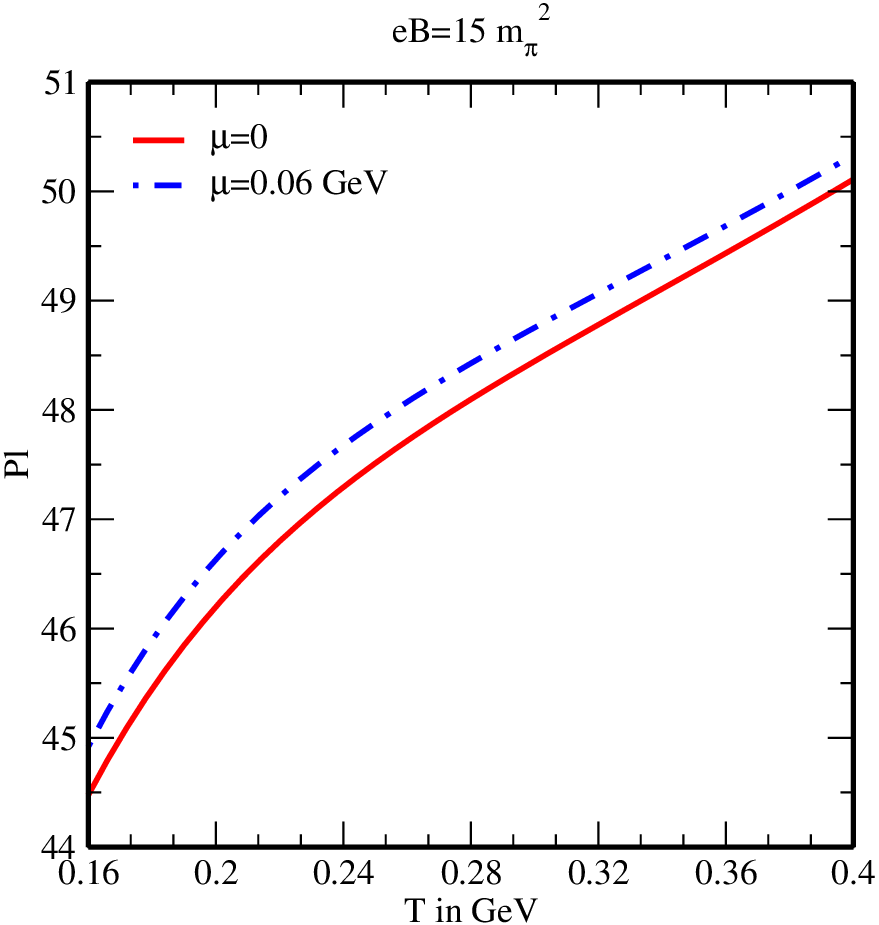} \\
a & b
\end{tabular}
\caption{Variations of the Prandtl number with temperature 
in the (a) absence and (b) presence of strong magnetic field.}\label{Fig.3}
\end{center}
\end{figure}

Figure \ref{Fig.3} depicts the variations of the Prandtl number (Pl) with the temperature in the absence and in the presence of strong magnetic field and chemical potential. We have observed that, Pl gets enhanced in the presence of a strong magnetic field and this value even gets larger in an additional presence of chemical potential. In all cases (figures \ref{Fig.3}a and \ref{Fig.3}b) Pl remains larger than unity, so, the momentum diffusion prevails over the thermal diffusion. Thus, the energy loss in a system due to the sound propagation is mainly governed by the momentum diffusion. The dominance of momentum diffusion over thermal diffusion is found to be more pronounced in the presence of both strong magnetic field and chemical potential. 

\subsection{Reynolds number}
\begin{figure}[]
\begin{center}
\begin{tabular}{c c}
\includegraphics[width=6.3cm]{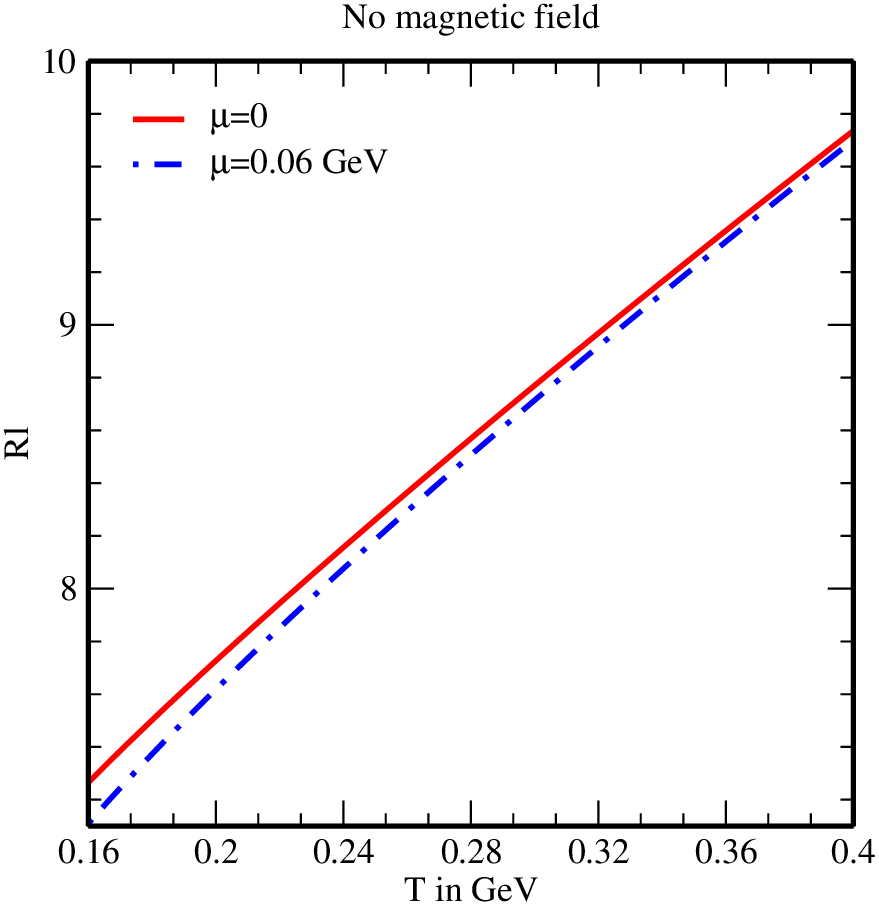}&
\hspace{0.74 cm}
\includegraphics[width=6.3cm]{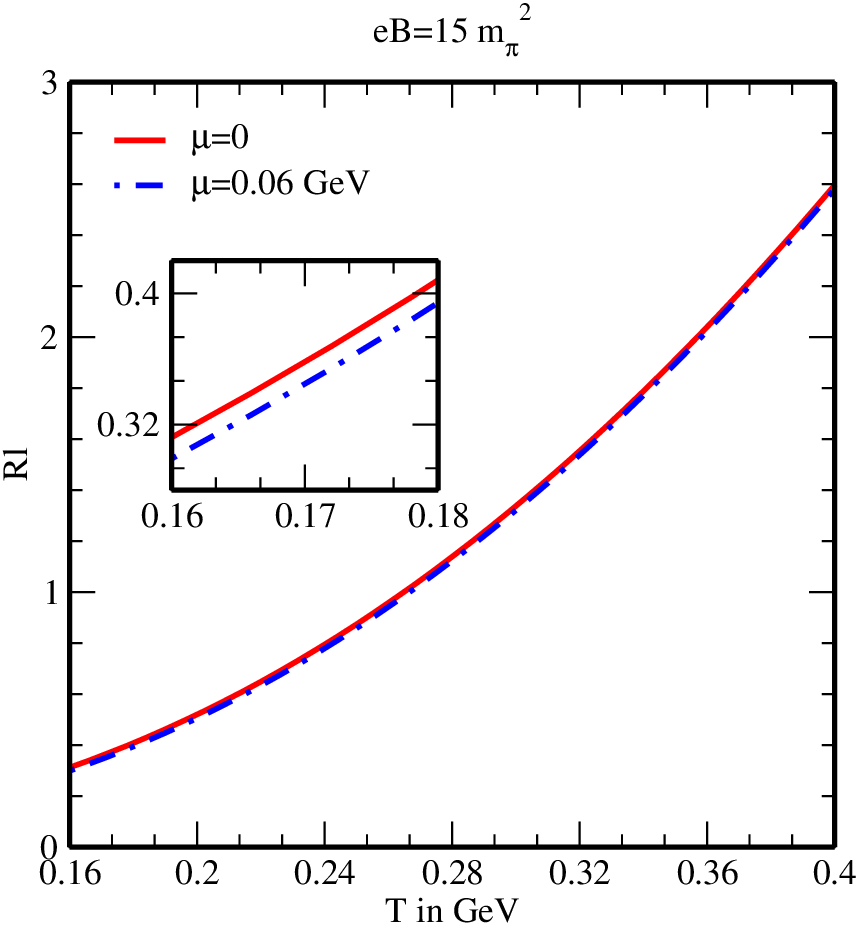} \\
a & b
\end{tabular}
\caption{Variations of the Reynolds number with temperature 
in the (a) absence and (b) presence of strong magnetic field.}\label{Fig.4}
\end{center}
\end{figure}

From figure \ref{Fig.4} we have observed that, the Reynolds number (Rl) increases with the increase of temperature irrespective of whether the strong magnetic field is absent or present in the medium, however, the presence of strong magnetic field does change the magnitude of the Reynolds number. It is noticed that the magnitude of Rl gets decreased in an ambience of strong magnetic field (figure \ref{Fig.4}b) as compared to that in the absence of magnetic field (figure \ref{Fig.4}a) and the existence of finite chemical potential further reduces its magnitude. The Reynolds number is even less than 1 at low temperatures up to $T\simeq0.26$ GeV (figure \ref{Fig.4}b), contrary to the zero magnetic field case where Rl remains larger than 1 over the entire range of temperature (figure \ref{Fig.4}a). Thus, for a dense QCD medium at strong magnetic field, the kinematic viscosity dominates over the characteristic length scale of the system and the flow is laminar, from which it is inferred that the presence of strong magnetic field and finite chemical potential makes the QCD medium more viscous. 

\subsection{Relative behavior between shear viscosity and bulk viscosity}
\begin{figure}[]
\begin{center}
\begin{tabular}{c c}
\includegraphics[width=6.3cm]{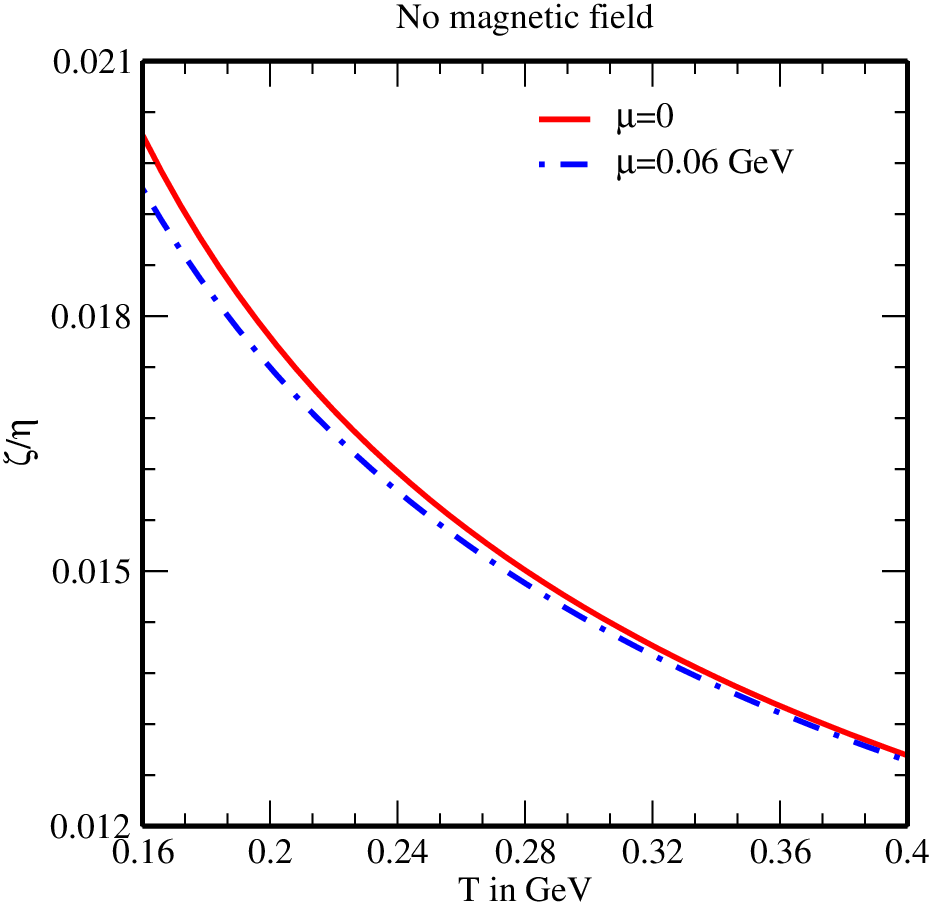}&
\hspace{0.74 cm}
\includegraphics[width=6.3cm]{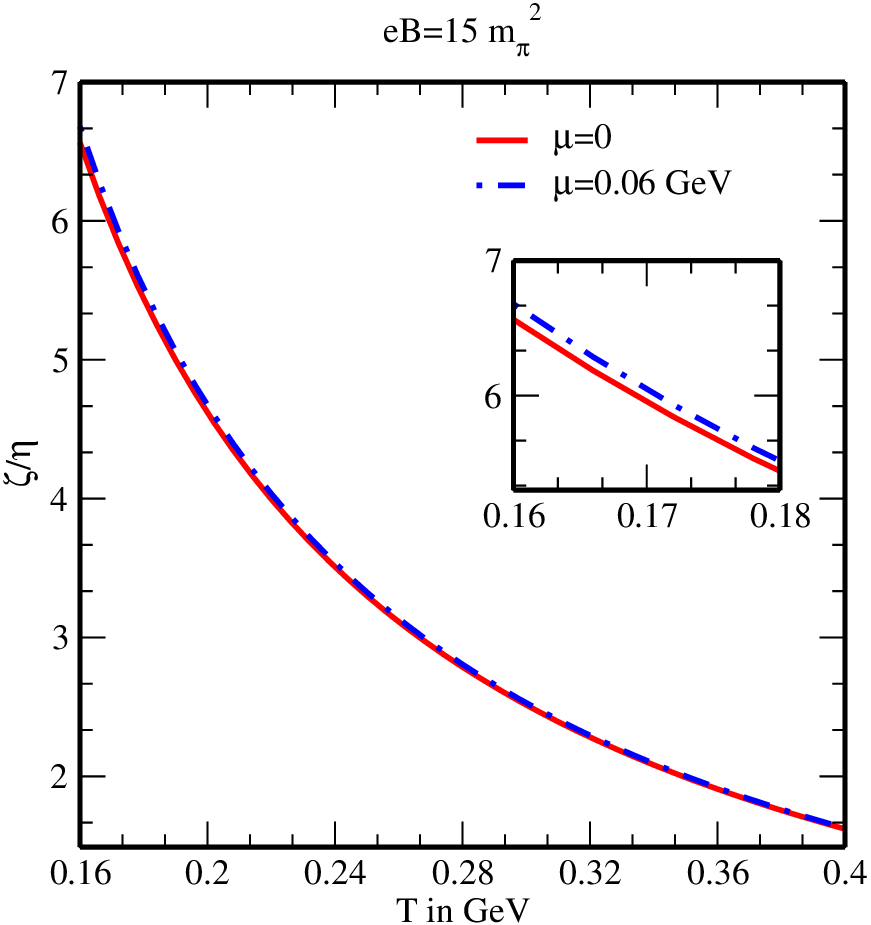} \\
a & b
\end{tabular}
\caption{Variations of $\zeta/\eta$ with temperature 
in the (a) absence and (b) presence of strong magnetic field.}\label{Ratio}
\end{center}
\end{figure}

Figure \ref{Ratio} shows how the ratio of the bulk viscosity to the shear viscosity ($\zeta/\eta$) varies with the temperature in the absence and in the presence of strong magnetic field and chemical potential. For the isotropic medium, $\zeta/\eta$ is much less than 1 (figure \ref{Ratio}a), whereas in the presence of a strong magnetic field this ratio surges and becomes greater than 1 (figure \ref{Ratio}b), so it explains that for an isotropic medium, the shear viscosity is larger than the bulk viscosity, whereas the opposite is true for a medium under the influence of a strong magnetic field. In all scenarios, the ratio $\zeta/\eta$ decreases with the increase of temperature. At finite chemical potential, $\zeta/\eta$ gets slightly increased in the strong magnetic field regime, contrary to the small decrease in the isotropic medium without the presence of magnetic field. So, the chemical potential helps in enhancing the dominance of shear viscosity over bulk viscosity at zero magnetic field, whereas it helps in enhancing the dominance of bulk viscosity over shear viscosity in an ambience of strong magnetic field. 

\section{Conclusions}
In this work, we have studied the viscous properties by calculating the shear and bulk viscosities of hot QCD matter in the presence of strong magnetic field and finite chemical potential and then explored some applications, such as the Prandtl number (Pl) to understand the interplay between momentum transport and heat transport properties, the Reynolds number (Rl) to know the relative behavior between viscous properties and characteristic length scale of the system, and the ratio $\zeta/\eta$ to interpret the competition between shear viscosity and bulk viscosity. We have determined the abovementioned quantities in the kinetic theory by solving the relativistic Boltzmann transport equation in the relaxation time approximation, where the interactions among particles are incorporated through their quasiparticle masses at finite temperature, strong magnetic field and finite chemical potential. Our observation has been carried out in the strong magnetic field limit, where the range of temperature and the value of chemical potential are kept much less than the strength of the magnetic field. 

Our studies on the shear and bulk viscosities shed light on how the strong magnetic field and finite but small chemical potential affect the viscous properties of hot QCD matter. We have observed that the presence of strong magnetic field lifts the values of $\eta$ and $\zeta$ as compared to the corresponding values in the isotropic medium at zero magnetic field, and these values get further increased in an additional presence of chemical potential. With the rising temperature, $\eta$ is found to increase, whereas $\zeta$ is found to decrease for the dense QCD medium under the influence of a strong magnetic field. The Prandtl number is observed to be greater than unity, so the momentum diffusion is larger than the thermal diffusion. However, when the medium is exposed to a strong magnetic field, Pl gets enhanced and it becomes further pushed to a higher value in the simultaneous presence of finite chemical potential. Thus, it explains that the energy dissipation due to the sound propagation in the medium is mostly regulated by the momentum diffusion. We have found a reduction in the Reynolds number in the abovementioned regime and it even becomes less than unity for temperatures below 0.26 GeV. So, it is inferred that the kinematic viscosity is larger than the characteristic length scale of the QCD medium, {\em i.e.} the QCD medium is more viscous in the presence of strong magnetic field and finite chemical potential than that in the absence of magnetic field and chemical potential, and the flow remains laminar. From our study on the ratio $\zeta/\eta$, it is observed that the shear viscosity prevails over the bulk viscosity for a dense QCD medium with no magnetic field, whereas the bulk viscosity dominates over the shear viscosity in the presence of a strong magnetic field. 

There exist some phenomenological implications of shear and bulk 
viscosities in heavy ion collisions. It is known that, the shear and 
bulk viscosities are associated with the transports of momentum across 
and along the layer, respectively. So, from their behaviors in strong 
magnetic fields, one can understand the effects of such fields on the 
momentum transport properties of the matter produced in the initial stages 
of heavy ion collisions. In case of the transition from hadronic phase to QGP 
phase, the values of shear and bulk viscosities are important in estimating the 
location of phase transition, where the shear viscosity attains a 
minimum value and the bulk viscosity attains a maximum value 
\cite{Csernai:PRL97'2006}. However, the emergence 
of strong magnetic field modifies the viscosities, thus, it could 
ultimately affect the phase transition of matter produced in the 
heavy ion collisions. In hydrodynamic simulations, the shear and 
bulk viscosities may also influence different observables, 
such as the elliptic flow coefficient, the hadron transverse 
momentum spectrum etc. \cite{Song:PLB658'2008,Denicol:JPG37'2010,Dusling:PRC85'2012,Noronha-Hostler:PRC90'2014}. So, the study of viscosities 
in the strong magnetic field regime can help in understanding how 
far the system appears from an ideal hydrodynamics in the said 
regime. With the help of the entropy density ($s$) of matter, the 
dimensionless ratios $\eta/s$ and $\zeta/s$ can be studied to 
get the information on how close the matter created at heavy ion 
collisions is to being perfect fluid and on the chiral symmetry, 
respectively, thus, our study on viscosities could be useful in 
interpreting the effect of strong magnetic field on such 
characteristics of matter. The observation on shear and bulk 
viscosities of the hot and dense QCD matter in an ambience of strong 
magnetic field also facilitates the understanding of the viscous 
properties in other areas where strong magnetic fields could be 
found, such as the cores of the dense magnetars and the beginning 
of the universe, apart from the ultrarelativistic heavy ion collisions. 

\section{Acknowledgment}
One of us (B. K. P.) is thankful to the Council of Scientific and 
Industrial Research (Grant No. 03(1407)/17/EMR-II) for the financial assistance. 

\appendix
\appendixpage
\addappheadtotoc
\begin{appendices}
\renewcommand{\theequation}{A.\arabic{equation}}
\section{Thermal conductivity in the presence of strong magnetic field and finite chemical potential}\label{A.T.C.}
For an isotropic dense QCD medium in the absence of magnetic field, thermal conductivity is given by
\be\label{I.T.C.}
\nonumber\kappa^{\rm iso} &=& \frac{\beta^2}{6\pi^2}\sum_fg_f\int d{\rm p} \frac{{\rm p}^4}{\omega_f^2} ~ \left[\tau_f(\omega_f-h_f)^2 ~ f_f^{\rm iso}(1-f_f^{\rm iso})\right. \\ && \left.+\tau_{\bar{f}}(\omega_f-\bar{h}_f)^2 ~ \bar{f}_f^{\rm iso}(1-\bar{f}_f^{\rm iso})\right]
.\ee
For a dense QCD medium in the presence of a strong magnetic field, thermal conductivity has the following form, 
\begin{eqnarray}\label{A.T.C.(eb)}
\nonumber\kappa^B &=& \frac{\beta^2}{4\pi^2}\sum_fg_f|q_fB|\int dp_3\frac{p_3^2}{\omega_f^2} ~ \left[\tau_f^B(\omega_f-h_f^B)^2f^B_f(1-f^B_f)\right. \\ && \left.+\tau_{\bar{f}}^B(\omega_f-\bar{h}_f^B)^2\bar{f}^B_f(1-\bar{f}^B_f)\right]
.\end{eqnarray}

\renewcommand{\theequation}{B.\arabic{equation}}
\section{Energy density and pressure in the presence of strong magnetic field and finite chemical potential}\label{A.(e,p)}
The energy density ($\varepsilon$) and the pressure ($P$) can be 
determined from the energy-momentum tensor ($T^{\mu\nu}$). For a 
medium in the absence of magnetic field, $\varepsilon$ and $P$ 
are respectively defined as
\begin{eqnarray}
\varepsilon &=& u_\mu T^{\mu\nu}u_\nu , \\
P &=& -\frac{1}{3}\left(g_{\mu\nu}-u_\mu u_\nu\right)T^{\mu\nu}
.\end{eqnarray}
In the presence of a strong magnetic field, the definitions 
of $\varepsilon$ and $P$ become modified as
\begin{eqnarray}
\varepsilon &=& u_\mu\tilde{T}^{\mu\nu}u_\nu , \\
P &=& -\left(g^\parallel_{\mu\nu}-u_\mu u_\nu\right)\tilde{T}^{\mu\nu}
.\end{eqnarray}

For a dense QCD medium in the absence of magnetic field, energy density 
is calculated as
\begin{eqnarray}\label{iso.(Er.D.)}
\varepsilon^{\rm iso} &=& \frac{1}{2\pi^2}\sum_f g_f \int d{\rm p}~{\rm p}^2\omega_f\left(f_f^{\rm iso}+\bar{f}_f^{\rm iso}\right)+\frac{1}{2\pi^2}g_g\int d{\rm p}~{\rm p}^2\omega_gf_g^{\rm iso}
,\end{eqnarray}
whereas, for a dense QCD medium in the presence of a strong magnetic field, 
energy density becomes
\begin{eqnarray}\label{baniso.(Er.D.)}
\varepsilon^B &=& \frac{1}{4\pi^2}\sum_f g_f|q_fB|\int d p_3\omega_f\left(f_f^B+\bar{f}_f^B\right)+\frac{1}{2\pi^2}g_g\int d{\rm p}~{\rm p}^2\omega_gf_g^{\rm iso}
.\end{eqnarray}

For a dense QCD medium in the absence of magnetic field, pressure 
is determined as
\begin{eqnarray}\label{iso.(P.)}
P^{\rm iso} &=& \frac{1}{6\pi^2}\sum_f g_f \int d{\rm p}~\frac{{\rm p}^4}{\omega_f}\left(f_f^{\rm iso}+\bar{f}_f^{\rm iso}\right)+\frac{1}{6\pi^2}g_g\int d{\rm p}~\frac{{\rm p}^4}{\omega_g}f_g^{\rm iso}
,\end{eqnarray}
whereas, in the presence of a strong magnetic field, pressure 
for a dense QCD medium becomes
\begin{eqnarray}\label{baniso.(P.)}
P^B &=& \frac{1}{4\pi^2}\sum_f g_f|q_fB|\int d p_3\frac{p_3^2}{\omega_f}\left(f_f^B+\bar{f}_f^B\right)+\frac{1}{6\pi^2}g_g\int d{\rm p}~\frac{{\rm p}^4}{\omega_g}f_g^{\rm iso}
.\end{eqnarray}

\renewcommand{\theequation}{C.\arabic{equation}}
\section{Thermal mass of quark in the presence of strong magnetic field and finite chemical potential}\label{Thermal mass (eB, T, C.P.)}
With the help of equations \eqref{q. propagator} and \eqref{g. propagator}, we have calculated the quark self-energy \eqref{Q.S.E.} in the imaginary time formalism at strong magnetic field and finite chemical potential, where the continuous energy integral ($\int\frac{dp_0}{2\pi}$) is replaced by the discrete Matsubara frequency sum and the integration over the transverse component of the momentum gives the factor $|q_fB|$. So the quark self-energy \eqref{Q.S.E.} takes the following form, 
\begin{eqnarray}\label{Q.S.E.(11)}
\nonumber\Sigma(p_\parallel) &=& \frac{2g^2}{3\pi^2}|q_fB|T\sum_n\int dk_z\frac{\left[\left(1+\gamma^0\gamma^3\gamma^5\right)\left(\gamma^0k_0
-\gamma^3k_z\right)-2m_f\right]}{\left[k_0^2-\omega^2_k\right]\left[(p_0-k_0)^2-\omega_{pk}^2\right]} \\ &=& \frac{2g^2|q_fB|}{3\pi^2}\int dk_z\left[(\gamma^0+\gamma^3\gamma^5)W^1-(\gamma^3+\gamma^0\gamma^5)k_zW^2\right]
,\end{eqnarray}
where $\omega^2_k=k_z^2+m_f^2$, $\omega_{pk}^2=(p_z-k_z)^2$ and the two frequency sums $W^1$ and $W^2$ are written as
\be
&&W^1=T\sum_n \frac{k_0}{\left[k_0^2-\omega_k^2\right]\left[(p_0-k_0)^2-\omega_{pk}^2\right]} ~, \\ &&W^2=T\sum_n \frac{1}{\left[k_0^2-\omega_k^2\right]\left[(p_0-k_0)^2-\omega_{pk}^2\right]}
~.\ee
After calculating the above frequency sums and then substituting, the quark self-energy \eqref{Q.S.E.(11)} can be simplified to take the following form, 
\begin{eqnarray}\label{Q.S.E.(2)}
\nonumber\Sigma(p_\parallel) &=& \frac{g^2|q_fB|}{3\pi^2}\int \frac{dk_z}{\omega_k}\left[\frac{1}{e^{\beta\omega_k}-1}+\frac{1}{2}\left\lbrace\frac{1}{e^{\beta(\omega_k+\mu_f)}+1}+\frac{1}{e^{\beta(\omega_k-\mu_f)}+1}\right\rbrace\right] \\ && \times\left[\frac{\gamma^0p_0+\gamma^3p_z}{p_\parallel^2}+\frac{\gamma^0\gamma^5p_z+\gamma^3\gamma^5p_0}{p_\parallel^2}\right]
.\end{eqnarray}
In the small chemical potential limit ($T>\mu_f$), we have evaluated the integration over $k_z$ and obtained the approximated result as
\begin{eqnarray}\label{Q.S.E.(3)}
\nonumber\Sigma(p_\parallel) &\approx& \frac{g^2|q_fB|}{3\pi^2}\left[\frac{\pi T}{2m_f}-\ln(2)+\frac{7\mu_f^2\zeta(3)}{8\pi^2T^2}-\frac{31\mu_f^4\zeta(5)}{32\pi^4T^4}\right] \\ && \times\left[\frac{\gamma^0p_0}{p_\parallel^2}+\frac{\gamma^3p_z}{p_\parallel^2}+\frac{\gamma^0\gamma^5p_z}{p_\parallel^2}+\frac{\gamma^3\gamma^5p_0}{p_\parallel^2}\right]
,\end{eqnarray}
where $\zeta(s)$ is the Riemann zeta function with $s=3,5$ here. 

For a thermal medium at finite magnetic field, the quark self-energy is written in the general covariant form \cite{Karmakar:PRD99'2019,Viscosities} as
\begin{equation}\label{general q.s.e.1}
\Sigma(p_\parallel)=A\gamma^\mu u_\mu+B\gamma^\mu b_\mu+C\gamma^5\gamma^\mu u_\mu+D\gamma^5\gamma^\mu b_\mu
~,\end{equation}
where $A$, $B$, $C$ and $D$ are the form factors, and $u^\mu$ (1,0,0,0) and $b^\mu$ (0,0,0,-1) are the directions of heat bath and magnetic field, respectively. The form factors are computed as
\begin{eqnarray}
A=\frac{1}{4}{\rm Tr}\left[\Sigma\gamma^\mu u_\mu\right]=\frac{g^2|q_fB|}{3\pi^2}\left[\frac{\pi T}{2m_f}-\ln(2)+\frac{7\mu_f^2\zeta(3)}{8\pi^2T^2}-\frac{31\mu_f^4\zeta(5)}{32\pi^4T^4}\right]\frac{p_0}{p_\parallel^2} ~, \\ 
B=-\frac{1}{4}{\rm Tr}\left[\Sigma\gamma^\mu b_\mu\right]=\frac{g^2|q_fB|}{3\pi^2}\left[\frac{\pi T}{2m_f}-\ln(2)+\frac{7\mu_f^2\zeta(3)}{8\pi^2T^2}-\frac{31\mu_f^4\zeta(5)}{32\pi^4T^4}\right]\frac{p_z}{p_\parallel^2} ~, \\ 
C=\frac{1}{4}{\rm Tr}\left[\gamma^5\Sigma\gamma^\mu u_\mu\right]=-\frac{g^2|q_fB|}{3\pi^2}\left[\frac{\pi T}{2m_f}-\ln(2)+\frac{7\mu_f^2\zeta(3)}{8\pi^2T^2}-\frac{31\mu_f^4\zeta(5)}{32\pi^4T^4}\right]\frac{p_z}{p_\parallel^2} ~, \\ 
D=-\frac{1}{4}{\rm Tr}\left[\gamma^5\Sigma\gamma^\mu b_\mu\right]=-\frac{g^2|q_fB|}{3\pi^2}\left[\frac{\pi T}{2m_f}-\ln(2)+\frac{7\mu_f^2\zeta(3)}{8\pi^2T^2}-\frac{31\mu_f^4\zeta(5)}{32\pi^4T^4}\right]\frac{p_0}{p_\parallel^2}
~.\end{eqnarray}
Here one can see that, $C=-B$ and $D=-A$. The quark self-energy \eqref{general q.s.e.1} can also be expressed in terms of the right-handed ($P_R=(1+\gamma^5)/2$) and left-handed ($P_L=(1-\gamma^5)/2$) chiral projection operators as
\begin{equation}\label{projection}
\Sigma(p_\parallel)=P_R\left[(A+C)\gamma^\mu u_\mu+(B+D)\gamma^\mu b_\mu
\right]P_L+P_L\left[(A-C)\gamma^\mu u_\mu+(B-D)\gamma^\mu b_\mu\right]P_R
~.\end{equation}
With the help of the substitutions $C=-B$ and $D=-A$, eq. \eqref{projection} turns out to be
\begin{equation}\label{projection1.1}
\Sigma(p_\parallel)=P_R\left[(A-B)\gamma^\mu u_\mu+(B-A)\gamma^\mu b_\mu
\right]P_L+P_L\left[(A+B)\gamma^\mu u_\mu+(B+A)\gamma^\mu b_\mu\right]P_R
~.\end{equation}
Substituting the quark self-energy \eqref{projection1.1} in Schwinger-Dyson equation, we get 
\be
\nonumber S^{-1}(p_\parallel) &=& \gamma^\mu p_{\parallel\mu}-\Sigma(p_\parallel) \\ &=& P_R\gamma^\mu X_\mu P_L+P_L\gamma^\mu Y_\mu P_R
~,\ee
where
\begin{eqnarray}
&&\gamma^\mu X_\mu=\gamma^\mu p_{\parallel\mu}-(A-B)\gamma^\mu u_\mu-(B-A)\gamma^\mu b_\mu ~, \\ 
&&\gamma^\mu Y_\mu=\gamma^\mu p_{\parallel\mu}-(A+B)\gamma^\mu u_\mu-(B+A)\gamma^\mu b_\mu
~.\end{eqnarray}
So, the effective quark propagator is obtained as
\be
S(p_\parallel)=\frac{1}{2}\left[P_R\frac{\gamma^\mu Y_\mu}{Y^2/2}P_L+
P_L\frac{\gamma^\mu X_\mu}{X^2/2}P_R\right]
,\ee
where
\begin{eqnarray}
&&\frac{X^2}{2}=X_1^2=\frac{1}{2}\left[p_0-(A-B)\right]^2-\frac{1}{2}\left[p_z+(B-A)\right]^2 ~, \\ 
&&\frac{Y^2}{2}=Y_1^2=\frac{1}{2}\left[p_0-(A+B)\right]^2-\frac{1}{2}\left[p_z+(B+A)\right]^2
~.\end{eqnarray}
Finally, the effective mass (squared) of quark in the presence of strong magnetic field, finite temperature and finite chemical potential is determined by taking the $p_0=0, p_z\rightarrow 0$ limit of either $X_1^2$ or $Y_1^2$ (both of them are equal in this limit) \cite{Rath:EPJC80'2020}, 
\begin{eqnarray}
m_{fT,B}^2=\frac{g^2|q_fB|}{3\pi^2}\left[\frac{\pi T}{2m_f}-\ln(2)+\frac{7\mu_f^2\zeta(3)}{8\pi^2T^2}-\frac{31\mu_f^4\zeta(5)}{32\pi^4T^4}\right]
.\end{eqnarray}

\end{appendices}

\end{document}